\documentclass{article}
\usepackage{aas_macros}
\usepackage{natbib}
\usepackage{graphicx}	
\usepackage[font=small]{caption}

\usepackage[T1]{fontenc}

\usepackage{hyperref}        
\usepackage{newtxtext}
\usepackage[varvw]{newtxmath}
\usepackage[dvipsnames]{xcolor}

\DeclareUnicodeCharacter{2212}{\textendash}

\begin{document}
\title{AGN Feedback in Groups and Clusters of Galaxies}
\author{Julie Hlavacek-Larrondo\thanks{Julie Hlavacek-Larrondo: Physics Department, Université de Montréal, Science Complex, CP 6128, succ. Centre-ville, Montréal, (Québec) H3C 3J7, Canada. email: j.larrondo@umontreal.ca}, Yuan Li\thanks{Yuan Li: University of North Texas, 1155 Union Circle \#311277 Denton, Texas 76203-5017. email: yuan.li@unt.edu}, Eugene Churazov\thanks{Eugene Churazov; Max Planck Institute for Astrophysics, Karl-Schwarzschild-Str. 1, D-85741 Garching, Germany; Space Research Institute (IKI), Profsoyuznaya 84/32, Moscow 117997, Russia. email: churazov@mpa-garching.mpg.de}}

%
%
\maketitle
\tableofcontents{}

\abstract{Active Galactic Nuclei (AGN) feedback stands for the dramatic impact that a supermassive black hole can make on its environment. It has become an essential element of models that describe the formation and evolution of baryons in massive virialized halos in the Universe. The baryons' radiative losses in the cores of these halos might lead to massive cooling of the gas and vigorous star formation on the order of $\sim10-1000\,$M$_\odot$ yr$^{-1}$, whereas observations show that the star formation rates are considerably less (i.e. $\sim1-10\,$M$_\odot$ yr$^{-1}$). It has now become clear from an observational, theoretical, and simulation perspective that the activity of central supermassive black holes  compensates for gas cooling losses and prevents very high star formation rates in massive galaxies, which otherwise would be much brighter than observed today. While AGN feedback is important over a broad range of halo masses, the most massive objects like galaxy groups and clusters truly provide outstanding laboratories for understanding the intrinsic details of AGN feedback. Partly, this is because in the nearby massive objects
we can directly see what AGN feedback is doing to its surrounding hot halo in exquisite details, as opposed to less massive or distant systems. 
} Yet another reason 
is that in the most massive objects the magnitude of AGN feedback has to be extremely large, providing the most stringent constraints on the models. 
In a nutshell, the AGN feedback paradigm in groups and clusters postulates that (i) a supermassive black hole in the center of a halo can release a vast amount of energy, (ii) this energy can be intercepted and thermalized by the gaseous atmosphere of the halo, and (iii) the system self-regulates so that the black hole energy release scales with the properties of the halo. A combination of multi-wavelength observations, in particular X-ray and radio wavebands, provides compelling evidence of the AGN feedback importance. Similarly, theoretical arguments suggest that self-regulation might be a natural property of a system consisting of the gaseous atmosphere and a black hole at the bottom of the potential well.

\section{Keywords} 
Clusters of galaxies; Groups of galaxies; Intracluster medium (ICM); active galactic nucleus feedback; supermassive black holes; accretion processes; galaxies.  




\section{Introduction}
\label{sec:intro}

Over the past two decades, AGN feedback has become an important and rapidly developing field. It comes in several flavors that reflect the mass of the host halo, the type of the gaseous AGN environment, the mass accretion rate, and the form of the AGN energy release. For instance, markedly different AGN feedback variants are associated with individual galaxies during the epoch of rapid star formation and black holes growth at $z\sim 2$ and in the present-day $z\sim 0$ massive systems like groups and clusters of galaxies. In the former, known as ``Quasar mode feedback'', AGN feedback can have the most violent form and is able to unbind the halo gas. 

\begin{figure}[!ht]
\includegraphics[width=0.98\columnwidth]{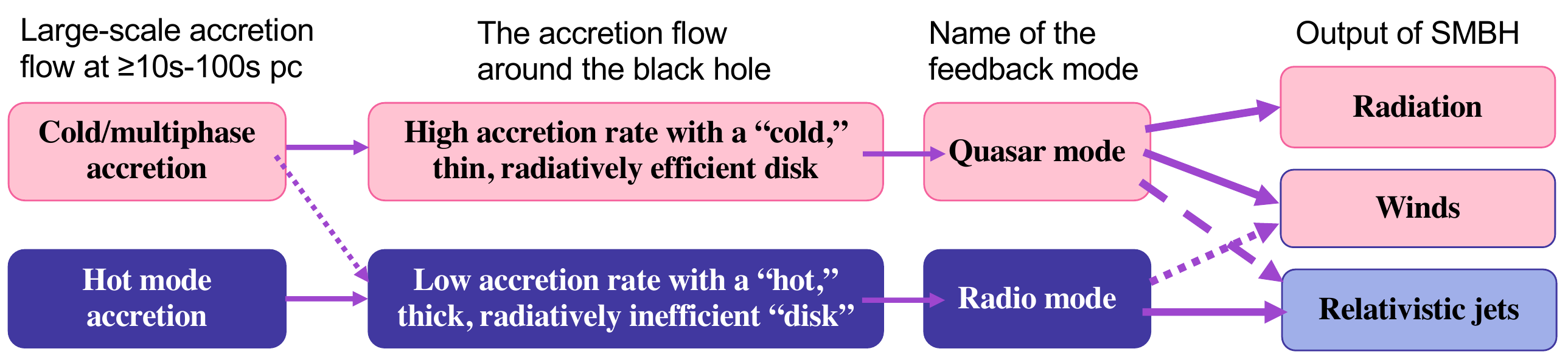}
\caption{The different AGN feedback modes and their connections to SMBH accretion. Quasar mode feedback is powered by a ``cold'', thin, radiatively efficient accretion disk around the SMBH. It occurs when the accretion rate is high and is likely fueled by multiphase gas from larger scales. The output of a quasar includes radiation and winds (likely driven by radiation). Some quasars also produce relativistic jets. Quasar mode feedback is often observed in galaxies at higher redshifts. Radio mode feedback is powered by a ``hot'', thick, radiatively inefficient ``puffy disk''. It may be fueled by hot mode Bondi accretion or multiphase gas from larger scales. The main output of radio mode feedback is the relativistic jets, comprised of relativistic particles that produce the observed radio emission. This radio mode is sometimes also called ``jet mode''. There may also be a non-relativistic wind produced along with the jets. Radio mode is often observed in low-redshift massive systems -- galaxy clusters, groups, and massive galaxies. It is thought to maintain the quiescent state of these systems and is therefore also called ``maintenance mode'' feedback. 
Radio mode feedback is sometimes also called kinetic or mechanical feedback mainly due to the way it is often modeled in numerical simulations, while quasar mode feedback is sometimes called thermal feedback as it is often modeled as thermal energy injection in simulations (see Section~\ref{sec:sim1} for details).}
\label{fig:feedback_modes}
\end{figure}

In the latter, a less violent form of the AGN feedback is observed, whose role is to keep the halo's gaseous atmosphere hot at present time, balancing the gas cooling losses rather than completely removing it from the halo. This second form of feedback is known as ``Radio mode feedback'' due to a widespread association of the feedback signatures with radio-bright bubbles inflated by AGN (see Figure~\ref{fig:feedback_modes} for a summary of different accretion and feedback modes discussed here). Thousands of papers have been written on observations, simulations, and theoretical analyses of this form of AGN feedback. Here we briefly outline some very general properties of this feedback mode and refer readers to extended reviews  \citep{2012ARA&A..50..455F,2012NJPh...14e5023M,2019SSRv..215....5W} for a detailed discussion relevant to galaxy groups and clusters.

The $\Lambda$CDM cosmological model makes clear predictions on the growth of small perturbations with time and their eventual transformation into virialized objects that decouple from the expansion of the Universe (see also the X-ray cluster cosmology Chapter). In particular, it can predict when objects more massive than a certain value start to form. At present time, the most massive are the objects, observationally known as clusters of galaxies, with a total mass $M_{\rm CL}\sim 10^{14-15}\, M_\odot$ \citep{2012ARA&A..50..353K}. Together with somewhat smaller groups of galaxies, they represent the high end of the virialized objects mass function in the present-day Universe.  On the mass scales corresponding to clusters, the amplitude of the primordial perturbations power spectrum is only moderately suppressed during the radiation-dominated phase of the Universe expansion. As a result, the gravitational potential wells of clusters are deep enough to assume that their baryon-to-total-mass ratio is similar to the cosmic baryon fraction. 

Yet, another property makes these massive objects special, namely the relation between the cluster characteristic dynamic time $t_{\rm dyn}$ and the time needed for infalling baryons to cool $t_{\rm cool}$ \citep{1977MNRAS.179..541R,1991ApJ...379...52W}. Here the dynamic time is defined as 

\begin{eqnarray}
t_{\rm dyn}=\frac{2R_\Delta}{v_\Delta},
\end{eqnarray}

where $R_\Delta$ is the cluster virial radius at a given overdensity $\Delta$ (relative to the critical density of the Universe) and the corresponding virial velocity

\begin{eqnarray}
v_\Delta=\left ( \frac{GM_\Delta}{R_\Delta}\right)^{1/2}.
\end{eqnarray}

With this definition (for $\Delta=200$), 
\begin{eqnarray}
t_{\rm dyn}=\frac{2}{\left ( \Delta/2\right )^{1/2}} t_{\rm H}=0.2 H(z)^{-1}  
\end{eqnarray}

and it is independent on the halo mass. In turn, $t_{\rm cool}$ is the ratio of the thermal energy density (or enthalpy) of the gas to its radiative losses, namely

\begin{eqnarray}
t_{\rm cool}=\frac{\gamma}{\gamma-1}\frac{nkT}{n^2\Lambda(T)},
\end{eqnarray}

where $\gamma=5/3$ is the gas adiabatic index for monoatomic non-relativistic gas, $n$ is the number density of the gas, and $\Lambda(T)$ is the cooling function of the gas that depends on temperature and metallicity \citep{1993ApJS...88..253S}. For massive clusters and groups, thermal bremsstrahlung makes the dominant contribution to $\Lambda(T)$. When $t_{\rm cool}$ is shorter than $t_{\rm dyn}$, baryons can cool during the infall, while in the opposite case a gaseous atmosphere is formed first. Since massive objects form later than the less massive ones, the characteristic density of the gas is lower in the former objects and the cooling time $\propto n^{-1}$ is also longer. As a result, both galaxy groups and clusters possess gaseous pressure-supported atmospheres with a gas temperature of the order of the virial temperature of the halo ($10^{7-8}\,{\rm K}$). It is exactly due to the presence of this gas that these objects become powerful X-ray sources and prominent peaks in the thermal Sunyaev-Zeldovich (SZ) effect maps \citep{1972CoASP...4..173S}. 

\begin{figure} 
\includegraphics[trim=0cm 5cm 0cm 3cm,width=1\columnwidth]{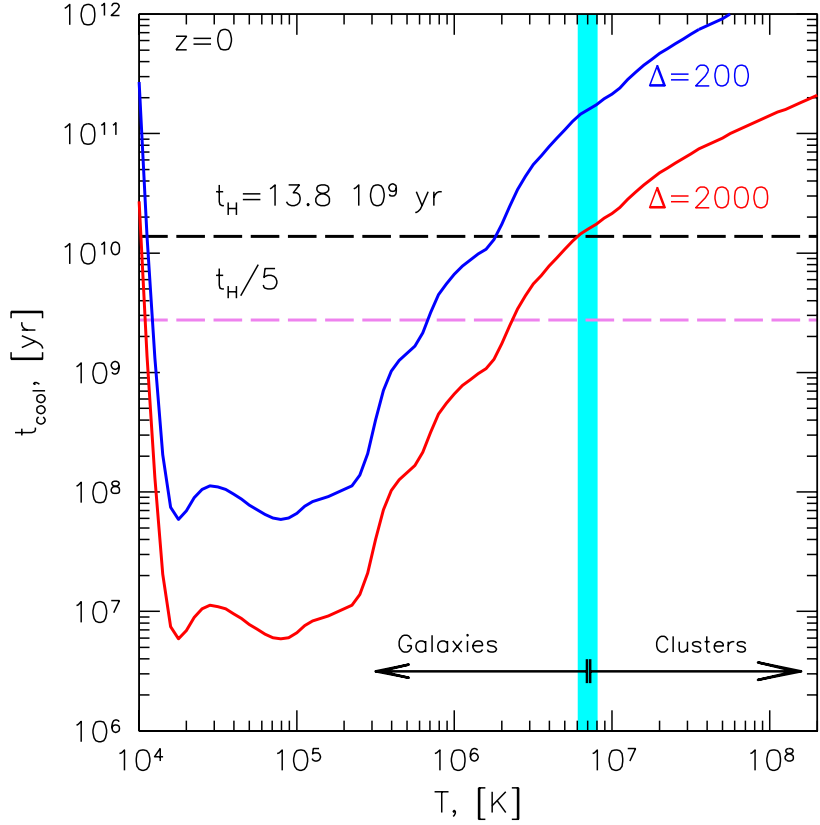}
\caption{Cooling time of the gas (baryons) as a function of the gas temperature at $z=0$. The (virial) temperature serves as a proxy to the mass of the system. The blue and red curves correspond to overdensities of the gas $\Delta=200$ and 2000, respectively, i.e. $\rho_{\rm b}=\Delta \Omega_{\rm b} \rho_{\rm c}$, where $\rho_{\rm c}$ is the critical density of the Universe.  A cooling function from \cite{1993ApJS...88..253S} with abundance of heavy elements 0.3 Solar is used. The metallicity of the gas has a big impact on the cooling time in the low-mass halos, but for groups and clusters, its impact is much less severe. Two dashed horizontal lines corresponds to the Hubble time $t_{\rm H}$ and dynamic time $t_{\rm dyn}\sim t_{\rm H}/5$, respectively. Below the $t_{\rm H}$ line, the gas cooling time is shorter than the age of the Universe. Below the $t_{\rm dyn}$ line, the cooling of the gas is important during the formation of the object and the gas cools before its forms a hydrostatic hot atmosphere.
The vertical cyan line shows schematically the division between the massive halos (galaxy groups and clusters) and the lower mass halos (galaxies). 
From the blue curve (overdensity $\Delta=200$) it is clear that for clusters and groups, (i) the cooling is not important for the formation of hydrostatic atmospheres and (ii) the bulk of the gas remains hot over the Hubble time. Only in the central regions (at overdensities larger than $\Delta=2000$, the red curve) this gas can cool. In these massive objects, the role of AGN feedback is to prevent the gas from cooling. } 
\label{fig:kt_tcool}
\end{figure}

The cooling time of the bulk of the gas in clusters and groups is longer not only than the dynamic time of individual objects but even than the Hubble time $t_{\rm H}=H_{\rm 0}^{-1}\approx 1.4\times10^{10}\,{\rm yr}$ at $z=0$. However, in the pressure-supported atmosphere at the virial temperature, the density in the central region of a halo is  much higher than the mean halo density and the corresponding cooling time is much shorter - down to $0.01-0.1\times t_{\rm H}$. 

This was recognized early on \citep{1973ApJ...184L.105L}, shortly after clusters were observed to be extended X-ray objects \citep{1971ApJ...167L..81G,1972ApJ...174L..65K,1972ApJ...178..309F} and their X-ray emission was interpreted as due to the thermal bremsstrahlung and lines of helium- and hydrogen-like ions of heavy elements like iron \citep{1976MNRAS.175P..29M}. Thus, it became clear that in the inner regions of cluster atmospheres (within the radius $r_{\rm cool}$ where the cooling time is equal to the Hubble time) the gas might cool, condense, and form new stars. Since the cooling time is the shortest in the core, the very central region will cool first, lowering the pressure and forcing outer gas layers to flow towards the center. This led to a concept of a ``cooling flow'', which was studied in great detail \citep{1994ARA&A..32..277F}. The observed high X-ray luminosity coming from cluster cores and the low gas temperature in the same regions served as strong arguments in favor of this concept. 

However, other observations, in particular, masses and colors of the central galaxies were in tension with the basic versions of the cooling flow model. Indeed, the observed luminosity can be straightforwardly converted into the mass rate of the gas cooling (also known as the mass deposition rate) such that

\begin{eqnarray}
\dot{M}\approx\frac{L_{\rm cool}}{kT}m_{\rm p}\approx 1600 \frac{L_{\rm cool}}{10^{45}\,{\rm erg\,s^{-1}}}\left(\frac{kT}{10\,{\rm keV}}\right)^{-1}\, M_\odot\,{\rm yr^{-1}},
\end{eqnarray}

where several factors of order unity have been omitted for clarity. If even a fraction of this mass is converted into stars over the Hubble time, this might lead to the accumulation of a huge stellar mass $M_*\sim 10^{13}\,M_\odot$ in the center of a cluster. While all relaxed clusters do host giant elliptical galaxies in their cores, these galaxies are at least an order of magnitude smaller. In addition, for the usual stellar mass function, the colors of these galaxies should be very blue due to the continuous formation of new stars, 
which is not observed (e.g. \cite{1999MNRAS.306..857C}). While various models have been suggested to mitigate this and other related problems in the general frame of the cooling flow concept, it was recognized that another possibility should be seriously considered, namely that there is a source of energy that compensates gas cooling losses, effectively preventing the cooling flow or, at least dramatically reducing the mass deposition rate. 

Various suggestions for this source of energy have been made, including thermal conduction, gas motions generated by mergers, cosmic rays, etc \cite[see][for review]{1994ARA&A..32..277F}. Currently, the leading hypothesis is that the mechanical energy of supermassive black holes, delivered by jets and/or winds/outflows is the most important source of energy for the cooling gas in groups and clusters. The term ``cool core'' is currently used instead of the ``cooling flow'' reflecting this paradigm shift.

Overall, studies indicate that (i) almost always the gas radiative losses are mitigated and (ii) if they are not mitigated (for whatever reason) the cluster can switch to a state resembling a ``cooling flow'' solution. There are several plausible reasons why these clusters are different from the vast majority of other clusters. The simplest one is that a recent merger has either brought a large amount of relatively cold gas to the system or displaced the cooling gas from the central black hole. If this is the case, these objects are in a transient state and will become more normal with time. In a much larger fraction of clusters, mergers apparently disperse the cooling gas in the core (the so-called ``non-cool-core clusters''). Those clusters should eventually return to the cool-core class too.  Yet another possibility is that in some clusters the black hole is not massive enough to produce a sufficient amount of energy in a mechanical form. In this case, the black hole can grow very rapidly and eventually become sufficiently massive to overcome the gas cooling and switch to the radio-mode feedback. 

One example in which cooling losses do not appear to be balanced by heating is the Phoenix cluster at redshift $z=0.597$ (see \cite{2012Natur.488..349M} and references therein). The Phoenix cluster is an extremely X-ray luminous cool core cluster that hosts a starburst associated with the central galaxy fueling $500-800{M_{\odot}}\,{\rm yr^{-1}}$ \citep{2012Natur.488..349M,2013ApJ...765L..37M}. The central galaxy also hosts an AGN that is incapable (for some reason) of preventing the entire hot halo from cooling (see \S\ref{BondiJulie}).
However, such objects are extremely rare, the fraction of clusters in a similar state does not exceed 1 per cent \citep{2021ApJ...910...60S}. 

While some clusters are heavily perturbed by ongoing mergers, many appear relatively relaxed and approximately spherically symmetric (see also Chapter on the dynamics of the X-ray emitting plasma in clusters). These systems contain very massive giant elliptical galaxies, which according to the bulge stellar mass - black hole mass $M_{\rm BH}$ relation (or the velocity dispersion - $M_{\rm BH}$ relation) should host extremely massive black holes $M_{\rm BH}\gtrsim 10^9\,M_\odot$. A crude estimate of the amount of energy one can, in principle, get from a massive black hole is

\begin{eqnarray}
E_{\rm BH}&=&\eta M_{\rm BH}c^2\sim 2\times10^{63}\,\left( \frac{\eta}{0.1}\right ) \left( \frac{M_{\rm BH}}{10^{10}\, M_\odot}\right ) \,{\rm erg}
\end{eqnarray}
where $\eta$ is a fraction of the accreted rest mass energy released by the black hole. For comparison, the total thermal energy of the gas in a massive cluster is 

\begin{eqnarray}
E_{\rm th}&=& \left( \frac{\Omega_{\rm b}}{\Omega_{\rm m}}\right ) \left( \frac{M_{\rm CL}}{\mu m_{\rm p}}\right ) \frac{1}{\gamma-1}kT \sim 7\times10^{63}\,\left( \frac{10^{15}}{M_\odot}\right ) \left( \frac{kT}{10\,{\rm keV}}\right )  \,{\rm erg},
\end{eqnarray}

Where $\frac{\Omega_{\rm b}}{\Omega_{\rm m}}$ is the mass fraction of baryons, $\mu\approx 0.61$ is the mean atomic weight per particle in cluster plasma, and $kT$ is the gas temperature.
The comparison of $E_{\rm BH}$ and $E_{\rm th}$ suggests that the total thermal energy of all baryons in a very massive cluster exceeds even the most optimistic estimates of the BH energetics. This is in contrast to the galaxy-mass halos, where the SMBH energy can unbind the entire baryonic content of the halo \citep{1998A&A...331L...1S}. 

In that case, more than enough energy is available from a black hole, whose mass obeys the well-established relation with the bulge mass or velocity dispersion, e.g. $M_{\rm BH}\approx 1.3 \times 10^8 \left ( \frac{\sigma_*}{200\,{\rm km\,s^{-1}}}\right)^{4.02}M_\odot$ \citep[][]{2002ApJ...574..740T}, and various solutions (even with a relatively low efficiency) might work. Here, $\sigma_*$ is the velocity dispersion of stars. A straightforward extrapolation of this relation to the richest clusters (replacing $\sigma_*$ with the velocity dispersion of galaxies $\sigma_{\rm gal} \sim 1000-1500 \,{\rm km\,s^{-1}}$) would imply $M_{\rm BH}\gtrsim 10^{11}\,M_\odot$, i.e. larger than found in the central galaxies in these clusters\footnote{This implies that if the black hole masses in those galaxies are also set by a variant of the AGN feedback model, then this model has lower efficiency than the one operating on cluster scales.}, although see \citet{2012MNRAS.424..224H,2018MNRAS.474.1342M} for evidence that some black holes in cluster cores may be over-massive. 

A different conclusion can be drawn from the analysis of the gas cooling losses within the sphere with radius $r_{\rm cool}$ and the energy release rate by SMBH over the Hubble time.  For a very luminous cluster

\begin{eqnarray}
L_{\rm cool}&\sim & 10^{45}\,{\rm erg\,s^{-1}} \\
L_{\rm BH}&=&E_{\rm BH}/t_{\rm H}\sim  4\times10^{45}\,\left( \frac{\eta}{0.1}\right ) \left( \frac{M_{\rm BH}}{10^{10}\, M_\odot}\right ) \,{\rm erg\,s^{-1}}.
\end{eqnarray}

In summary, $L_{\rm BH}$ can be comparable or larger\footnote{One can further reduce the power requirements (by a factor of 2-3) by noting that once in a while clusters experience significant mergers that can  disperse and  reheat cooling gas in the cluster core. Therefore the energy input from the SMBH is needed only in a fraction of the Hubble time, that is - between successive mergers.} than $L_{\rm cool}$.

Another useful quantity to consider is the ratio of $L_{\rm cool}$ to the Eddington luminosity of the BH 
\begin{eqnarray}
R_{\rm Edd}=\frac{L_{\rm cool}}{L_{\rm Edd}}\sim 10^{-3} \left ( \frac{L_{\rm cool}}{10^{45}\,{\rm erg\,s^{-1}}}\right ) \left (\frac{M_{\rm BH}}{10^{10}M_\odot} \right )^{-1},
\label{eq:redd}
\end{eqnarray}
where the Eddington luminosity $L_{\rm Edd}$ is set by the condition that the radiation pressure exerted by a compact source on electrons via Thomson scattering is balanced by the gravitational attraction acting on protons, i.e., $\displaystyle L_{\rm Edd}=\frac{4\pi c Gm_{\rm p}}{\sigma_{\rm T}} M_{\rm BH}$. A related quantity is the Eddington accretion rate $\dot{M}_{\rm Edd}=L_{\rm Edd}/\varepsilon$, where $\varepsilon$ is the radiative efficiency of the accretion flow (see eq.~\ref{eq:mdedd} below).
For the relevant range of cooling luminosities and black hole masses, $R_{\rm Edd}\sim 10^{-4}-10^{-2}$.

From the above discussion, which uses only the very basic observables and assumptions, one could draw three conclusions:
\begin{enumerate}
\item Massive black holes are not powerful enough to affect the bulk of the gas in a very rich cluster, but they can offset the gas cooling losses in the cluster core and, therefore, can, in principle, maintain the gas in a quasi-steady state. 
\item The efficiency of converting the rest mass of accreted matter into gas heating must be very high. If less than $\sim$10\% of the energy released by AGN goes into gas heating then extremely massive black holes would be needed in the cores of clusters. 
\item If an AGN operates continuously, then it might be sufficient to accrete matter at a level of $10^{-4}-10^{-2}$ of the Eddington rate to release an amount of energy comparable to gas cooling losses (provided that this energy efficiently couples to the gas).   
\end{enumerate}

Thus, groups and clusters of galaxies represent a class of the most massive virialized objects in the present-day Universe. Due to their large mass and, consequently, relatively late formation epoch, they possess hot pressure-supported atmospheres. However, the cores of these atmospheres are characterized by a short cooling time and without an external source of energy the gas should cool. 

Central supermassive black holes in these halos are plausible sources of energy that can compensate for gas cooling losses and prevent massive star-bursts, although the process of heating the must be very efficient. Many such objects have been extensively observed in all segments of the electromagnetic spectrum, offering a possibility to study the AGN feedback in great detail in the limiting case of the most powerful sources and the most ``clean'' and simple environment.



\section{Observational signatures of AGN feedback}
\label{sec:obs}
\subsection{Historical perspective}\label{obshistory}

\textit{Uhuru}, the first X-ray based satellite, launched in 1970, led to the discovery of the X-ray emission in clusters of galaxies \citep{1971ApJ...167L..81G,1971ApJ...165L..49K,1972ApJ...178..309F}. In the late 1970’s and early 1980’s, NASA launched three new satellites called the High Energy Astronomy Observatories (\textit{HEAO}), including the first X-ray imaging satellite named \textit{Einstein}. 
It was the Einstein observatory, which imaged many clusters in X-rays \citep{1984ApJ...276...38J}  and first detected disturbances in the intracluster plasma of the Perseus cluster \citep[i.e. the depression in X-ray surface brightness North-West off the nucleus][]{1981ApJ...248...55B,1981ApJ...248...47F}, although these were not yet recognized as related to AGN activity. Some ten years later, the \textit{ROSAT} High Resolution Imager (HRI) was able to resolve the inner X-ray cavities in Perseus (also known in the literature as ``radio bubbles'' or simply ``bubbles''). In combination with VLA radio data \citet{1990MNRAS.246..477P}, these structures were recognized as radio-bright bubbles that have been inflated by the radio jets of the central AGN \citep{1993MNRAS.264L..25B}. As we will see throughout this Chapter, a combination of the total energy (enthalpy) estimates of these bubbles with their ages set by buoyancy, led to the conclusion \citep{2000A&A...356..788C} that the AGN power in the Perseus cluster is comparable to the gas cooling losses. It was further conjectured that the North-West depression seen in X-ray images is a “ghost bubble” related to previous episodes of the AGN activity, which became dim in the radio band due to aging of relativistic electrons. 

Yet, it was the advent of the \textit{Chandra} X-ray observatory that revolutionized the field of AGN feedback, with its unprecedented subarcsecond resolution (see the Chapter dedicated to the \textit{Chandra} X-ray Observatory in Section III). In fact, \textit{Chandra} signified an order of magnitude improvement in the spatial resolution of X-ray imaging, and remains until now unrivaled in terms of high precision imaging, with a point spread function\footnote{The PSF measures the percentage of the total energy of a point source that falls within a given radius.} (PSF) of only $\sim1.0''$. 


Hydra A was the first nearby cluster of galaxies imaged with \textit{Chandra}. Previous X-ray telescopes, especially the \textit{Einstein} and \textit{ROSAT} HRI telescopes, clearly showed that galaxy clusters contained huge reservoirs of X-rays gas, hinted to some structure, but it was only when \textit{Chandra} was launched that it revealed in its full complexity and richness, the structure of the hot gas in clusters. Hydra A was observed on 9 December 1999 during its orbital verification and activation phase for a total integration time of 40 ks. These images revealed two disturbances in the intracluster gas: both centered around the central AGN and located in symmetry between one another \citep[][see Figure \ref{fig:obsfeedback})]{2000ApJ...534L.135M}. It was realized that similar to the Perseus cluster these structures were in fact large radio bubbles (X-ray cavities) that have been inflated by the relativistic jets originating from the AGN in the central cluster galaxy\footnote{Also known as the central dominant galaxy or the brightest cluster galaxy.} 


Many more examples of bubbles/X-ray cavities were soon found with \textit{Chandra} and \textit{XMM-Newton} proving that a similar process is happening in halos ranging from individual elliptical galaxies to the most luminous clusters. Furthermore, the estimated mechanical power of the jets turned out to be correlated with the luminosity of the cluster (see Section \ref{obsenergetics} below). It became obvious that AGN can, and plausibly does, affect the thermal state of the cooling gas in clusters and groups. Furthermore, it was realized that feedback from supermassive black holes may be a fundamental and essential process in the formation and evolution of all galaxies.






\subsection{The case of AGN feedback in groups and clusters of galaxies}\label{obscase}

After Hydra A was first imaged by \textit{Chandra}, astronomers seeking to determine how black hole feedback works started to target other black holes at the centres of galaxy groups and clusters \citep[see ][for reviews]{2007ARA&A..45..117M,2012NJPh...14e5023M}.

As argued above, at extreme temperatures of tens of millions of degrees, the hot intracluster gas was expected to cool down to temperatures of $\sim$30 Kelvin in less than a few hundred million years \citep{1994ARA&A..32..277F}. Once cooled, this gas should have deposited itself onto the central cluster galaxy, and extreme star formation rates of hundreds to thousands of solar masses per year were expected. However, spectroscopic studies of this gas, combined with a search for the cold gas and the expected newly-formed stars, revealed that it was not cooling at the predicted rates - it was only able to cool to 10 per cent or less of the predicted rates \citep[e.g. ][]{1999MNRAS.306..857C,2007MNRAS.379..867V}. Indeed, observations of the Fe L complex using the \textit{Chandra} and the \textit{XMM-Newton} space telescopes, which allow precise measurements of gas cooling at temperatures between $0.4$ and $4.0$ keV, showed that the intracluster gas appeared to cool down to a third of the virial temperature, but not further below \citep[see review by][]{2006PhR...427....1P}.
This was further supported by high-quality data from the Cosmic Origins Spectrograph onboard the Hubble Space Telescope, which found that the $\sim10^{5.5}$K gas, probed by the O VI emission line in the far UV, is cooling, but at rates an order of magnitude smaller than the predicted rates \citet{2014ApJ...791L..30M,2017ApJ...835..216D}. 

\begin{figure} 
\includegraphics[trim=0cm 0cm 0cm 0cm,width=1\columnwidth]{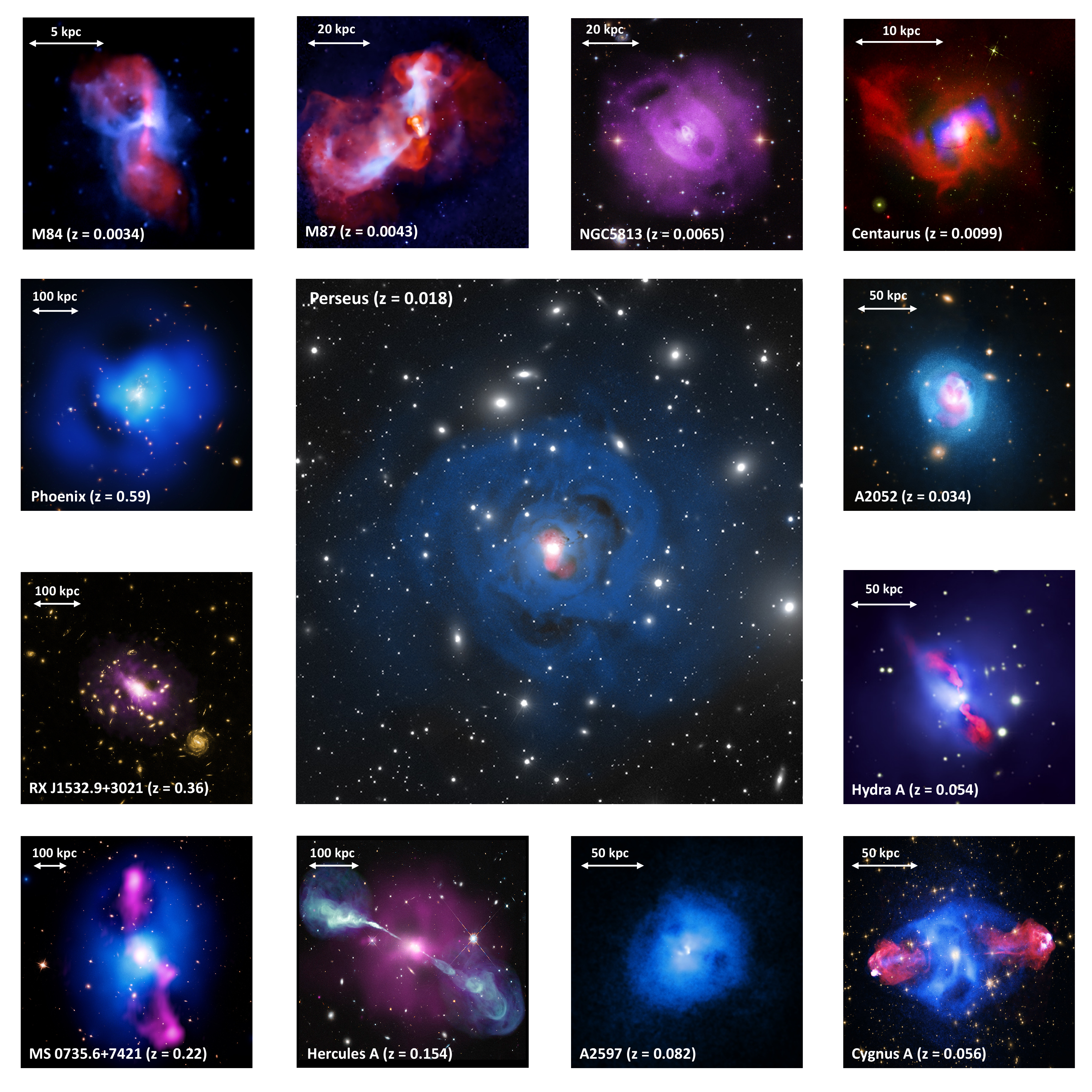}
\caption{\small{Collage of systems in which radio mode AGN feedback is taking place. The central image illustrates the Perseus cluster of galaxies, a prototypical example of radio mode AGN feedback. Here, X-rays are shown in blue (NASA/CXC), radio is shown in pink (NSF/NRAO/VLA) and the background image illustrates the galaxies (SDSS). Credit from MLGM/JHL/MPL. The images surrounding the central image are in order of redshift, starting from the top left corner and moving clockwise. For M84, X-rays are in blue (NASA/CXC/MPE/A.Finoguenov et al.), radio in pink (NSF/NRAO/VLA/ESO/R.A.Laing et al) and optical in yellow (SDSS), see \citet{2008ApJ...686..911F}. For M87, X-rays are in blue (NASA/CXC/KIPAC/N. Werner, E. Million et al) and radio in pink (NRAO/AUI/NSF/F. Owen), see 
\citet{2010MNRAS.407.2063W}. For NGC 5813,  X-rays are in purple (NASA/CXC/SAO/S.Randall et al.) and optical in yellow (SDSS), see \citet{2015ApJ...805..112R}. For Centaurus, X-rays are in red (NASA/CXC/MPE/J.Sanders et al.), optical in yellow (NASA/STScI) and radio in blue (NSF/NRAO/VLA), see \citet{2016MNRAS.457...82S}.
For Abell 2052, X-rays are in blue (NASA/CXC/BU/E.Blanton), radio is pink (NRAO/AUI/NSF) and optical in yellow (ESO/VLT), see \citet{2011ApJ...737...99B}. For Hydra A, X-rays are in blue (NASA/CXC/U.Waterloo/C.Kirkpatrick et al.) and radio is in pink (NSF/NRAO/VLA) and optical in yellow (Canada-France-Hawaii-Telescope/DSS), see \citet{2009ApJ...707L..69K}. For Cygnus A, X-rays are in blue (NASA/CXC/SAO),  optical is in yellow (NASA/STScI) and radio in pink (NSF/NRAO/AUI/VLA). For Abell 2597, X-rays are in blue (NASA/CXC/Michigan State Univ/G.Voit et al), optical in yellow (NASA/STScI \& DSS) and ionized gas in pink (Carnegie Obs./Magellan/W.Baade Telescope/U.Maryland/M.McDonald), see \citet{2018ApJ...865...13T}. For Hercules A, X-rays are in purple (NASA/CXC/SAO), optical in white (NASA/STScI),  and radio in blue (NSF/NRAO/VLA), see \citet{2005ApJ...625L...9N}. For MS 0735.6+7421, X-rays are in blue (NASA/CXC/SAO) and radio in pink (NSF/NRAO/VLA), see \citet{2005Natur.433...45M}. For RX J153.9+3021, X-rays are in blue (NASA/CXC/Stanford/J.Hlavacek-Larrondo), optical in yellow ( NASA/ESA/STScI/M.Postman \& CLASH team), see \citet{2013ApJ...777..163H}. For Phoenix, X-rays are in blue (NASA/CXC/MIT/M.McDonald et al), optical in yellow (NASA/STScI) and radio in pink (TIFR/GMRT), see \citet{2015ApJ...811..111M}}.}
\label{fig:obsfeedback}
\end{figure}

The accumulation of countless observations has made a clear case that it is the supermassive black hole in the central cluster galaxy that is preventing this gas from cooling and forming stars by driving powerful relativistic jets that carve out X-ray cavities. Indeed, by now, dozens of spectacular examples of AGN-driven jetted outflows have been discovered across the entire mass scale of galaxies. This is shown in Figure \ref{fig:obsfeedback}, where we highlight some of the most outstanding examples discovered to date. Note that X-ray cavities range roughly between 20 and 200 kpc in size, i.e. they typically extend well beyond the central cluster galaxy. 

In fact, the primary reason why AGN feedback can be studied in such detail in groups and clusters (as opposed to isolated galaxies), is that we can see the impact radio jets have on their surrounding medium. Indeed, we can literally see them carve out the cavities in the hot X-ray bright intracluster gas. In more isolated systems (and usually less massive galaxies), the gas that surrounds these galaxies is usually not bright enough to be visible with today's X-ray telescopes. In those galaxies, it therefore becomes extremely difficult to understand what radio jets are doing to their surroundings - this does not mean that they are not doing any work (and hence any impact on their surroundings), it just implies that feedback processes may be more difficult to see and understand in more isolated systems. Groups and clusters are therefore prime targets that allow astronomers to understand how AGN feedback works and the role supermassive black holes play in the formation and evolution of massive galaxies. 
 
Another aspect, related to the visual imprints of AGN feedback in groups and clusters, is that these outflows provide clean estimates for the energetics associated with black hole feedback; this is determined by straightforward calculations of the energy required to inflate the X-ray cavities (see Section \ref{obsenergetics}). Early studies based on X-ray cavities in groups and clusters clearly demonstrated that X-ray cavities are common in samples of groups and clusters that require some form of heating to prevent the gas from cooling \citep[e.g. ][]{2006ApJ...652..216R}, and that the energetics associated with these outflows are colossal \citep[e.g. ][]{2005Natur.433...45M}. 


\subsection{How does AGN feedback work (from an observational perspective)}\label{obswork}

Now that we have made the case of AGN feedback in groups and clusters, the questions that remain are how does it all work? What do these radio jets actually do to their environment, besides pushing away the hot X-ray gas and carving out X-ray cavities? If one inflates a balloon in a room, the room does not heat up. So besides the mechanical work it takes to inflate the bubbles, how is the energy then propagated throughout the group and cluster? Is it sufficient to explain why the hot X-ray halos are not cooling, and what is the duty cycle of such outflows? That is, how often is the AGN actively inflating X-ray cavities? And are they only occurring in modern times, or were they in place when the first groups and clusters were created? In the next section, we will explore all of these questions in detail. Giant leaps have been accomplished since the first 
detection of clear signatures of SMBHs impact on the ambient gas, and 
we now understand that AGN feedback (via radio jets) has a vital and essential role in all massive galaxies. 

\subsubsection{Accretion processes and modes}\label{BondiJulie}

The first question to address is how are the black holes fed, and similarly what are the typical accretion rates of these objects. 

What is clear from observations is that most AGN in groups and clusters of galaxies that have radio jets are not X-ray bright. In fact, most are radiatively inefficient and accreting at rates well below the Eddington rate, typically well below $\sim0.01\dot{M}_{\rm Edd}$. In this case, it is said that these AGN are in the radiatively inefficient state and that most of the energy that they release is in the form of radio jets \citep[see also][for other terms used to distinguish between the different modes, including 'high-excitation' ('quasar-mode') and 'low-excitation' ('radio-mode') radio galaxies]{2006MNRAS.368L..67B,2012MNRAS.421.1569B,2007MNRAS.376.1849H}. A summary of AGN feedback modes and their relation to accretion is also illustrated in Figure~\ref{fig:feedback_modes}.

Observationally, this comes mostly from our understanding of stellar mass black holes in which we see (on human time scales) black holes transition between the radiatively-efficient ($>0.01\dot{M}_{\rm Edd}$) to the radiatively-inefficient mode ($<0.01\dot{M}_{\rm Edd}$), see for example the evolution of GX 339-4 \citep{2005A&A...440..207B,2003A&A...400.1007C}. In the case of stellar mass black holes, we often refer to the different states as the soft and hard state (as well as a multitude of intermediate states). As the mass accretion rate changes, the physics governing the accretion disk also changes. In Section \ref{accretionEugene}, we explore this topic from a theoretical perspective in more detail.

In the case of galaxy clusters, while most of the central AGN are in the radiatively inefficient state, there are some examples of central AGN that are X-ray bright. This includes IRAS 09104+4109 \citep[e.g.][]{2012MNRAS.424.2971O}, H1821+643 \citep[e.g.][]{2010MNRAS.402.1561R} and 3C186 \citep[e.g.][]{2010ApJ...722..102S}), see also Figure 12 from \citet[][]{2013MNRAS.432..530R}, original plot from \citep{2005MNRAS.363L..91C}. 

In the three examples mentioned, the central AGN appear to be consistent with being in the radiatively efficient state, accreting at rates above $\sim0.01\dot{M}_{\rm Edd}$. They are therefore said to be in quasar mode feedback, where most of the energy released is in the form of radiation and winds (see Figure~\ref{fig:feedback_modes}). However, radiation pressure can not have an important impact on the scale of the galaxy - instead, the interaction is thought to be via dust (because the dust cross-section is much higher and dust grains will be partially charged near the quasar, which will bind them to the surrounding partially-ionized gas) or winds \citep[e.g. ][]{2011ApJ...729L..27R}. For the latter, winds speeds above 10,000 km s$^{-1}$ are commonly found via line absorption (i.e. the quasar continuum emission is absorbed by intervening wind material and gets blue-shifted). The energy produced 
by the action of AGN radiation pressure
is thought to be eventually deposited into the interstellar medium. It then compresses this medium into a shell that then blows the cold gas reservoir out of its host galaxy. Quasar mode feedback is commonly invoked to be important at earlier epoch \citep[e.g. ][]{2006MNRAS.365...11C}. 

However, it is important to note that all three of the examples mentioned also appear to exhibit evidence of radio jets. This implies that in some cases, quasar mode black holes can also form radio jets. 

A fourth spectacular example of such a black hole is the central AGN in the Phoenix cluster of galaxies mentioned earlier in this Chapter. The Phoenix cluster was discovered by the South Pole Telescope \citep[SPT-CLJ2344-4243;][]{2011PASP..123..568C}. It is an extremely X-ray luminous cool core cluster that hosts a starburst associated with the central galaxy with a star formation rate of $500-800{M_{\odot}}yr^{-1}$ \citep[][and references therein]{2012Natur.488..349M}. The central AGN shows evidence of extremely powerful radio jets \citep[e.g. ][]{2015ApJ...805...35H}, yet it is also a quasar. In fact, even though there is evidence of powerful radiative and mechanical AGN feedback processes, the feedback appears to be incapable of preventing the entire hot X-ray halo from cooling, leading to the starburst in the core.

One note is that all of the sources, including Phoenix, appear to be at high-redshift. \citet{2013MNRAS.431.1638H} argue that we are witnessing the transition between radio mode to quasar mode at high redshift. This is not surprising considering that the average accretion rate of black holes increases with redshift and peaks at about $z=2$ \citep[e.g. ][]{2007ApJ...654..731H}. Overall, central cluster AGN in quasar mode feedback appear to be extremely rare \citep[although this might be affected by selection biases or evolution processes, e.g. ][]{2021ApJ...910...60S}.  

So the question that remains is how are the black holes fed. As mentioned earlier, in Section \ref{accretionEugene}, we address this question from a theoretical perspective and present the two main modes of accretion discussed in the literature, essentially the ``hot'' and ``cold'' mode.

Initially, it was thought that the black holes at the centers of groups and clusters of galaxies accreted via the ``hot'' mode, also known as Bondi accretion. Such accretion relies on many assumptions including adiabaticity, spherical symmetry, unperturbed and steady initial conditions, as well as an absence of magnetic fields, self-gravity and feedback processes. If this is the case, then the advantage is that the nonlinear hydrodynamic equations can be solved analytically \citep{1952MNRAS.112..195B}. We invite the reader to see Section \ref{sec:models} for more details on why such a model works well for connecting how the black hole is fed to the larger scales, as well as Section \ref{accretionYuan} for how simulations incorporate such accretion. 

Observationally, there are some studies that are in favor of this ``hot'' accretion mode. In particular, \citet{2006MNRAS.372...21A} initially showed that for the very nearby systems with X-ray cavities, for which the Bondi radius is resolvable, the X-ray cavity power seemed to correlate with the Bondi accretion rate. Hence, this strongly implied that the accretion process of the central black holes in groups and clusters of galaxies seemed to be connected to Bondi accretion. However, this work was then revised by \citet{2013MNRAS.432..530R}, who found no evidence of a correlation between the X-ray cavity power and the Bondi accretion rate. The main difference between these two pieces of work relied on including more objects and using deeper observations for some objects. There have also been other studies that seem to suggest, at least on an observational side, that the black holes are not fed by Bondi accretion. This includes the mass accretion rate in M87 that appears to be at least 100 times less than the Bondi rate, in agreement with Faraday rotation measurements \citep{2015MNRAS.451..588R}. This also seems to be the case for SgrA* \citep{2003ApJ...591..891B} and NGC 3115 \citep{2014ApJ...780....9W}. Bondi accretion may also not work for high power jets \citep{2011ApJ...727...39M}.

More recently, another model has gained interest in the community, namely chaotic cold accretion \citep[e.g. ][]{2005ApJ...632..821P,2013MNRAS.432.3401G}. We explore this model in more detail in Section \ref{sec:models}. Observationally, this has also gained popularity because of studies that have found observational evidence of cold, clumpy accretion onto a black hole \citep[][]{2016Natur.534..218T}.

\subsubsection{Energetics and timescales}\label{obsenergetics}

\citet{2004ApJ...607..800B} and \citet{2006ApJ...652..216R} initially found that around 25 per cent of clusters of galaxies, including cool and non cool core clusters, seem to harbor X-ray cavities. However, X-ray cavities are predominantly found in cool core clusters of galaxies \citep[see also work on simulations such as by][]{2015ApJ...813L..17R}. Out of the 55 brightest clusters in the sky, \citet{2006MNRAS.373..959D,2008MNRAS.385..757D} found that 70 per cent of those requiring heating appeared to harbor clear X-ray cavities. Updating their work, we now know that 95 per cent of the 55 brightest clusters requiring heating (i.e. 19 of these 20 clusters) have X-ray cavities, and that all harbor a central radio source \citep{2012ARA&A..50..455F}. The only exception which lacked X-ray cavities was the Ophiuchus cluster. However, a recent study found tentative evidence of an extremely large X-ray cavity, although deeper observations are needed to confirm this result \citep{2020ApJ...891....1G}. Overall, this implies that the time in which the AGN is actively driving radio jets and inflating X-ray cavities, known as the duty cycle, is extremely high in cool core clusters and may even reach 100 per cent. 

Of course, there will be many biases that will affect this calculation. The primary one will most likely be the quality of the data \citep[see work by ][]{2014MNRAS.444.1236P}. This will limit the number of clusters in which we can identify X-ray cavities, and the number of X-ray cavities should be viewed as a lower limit to the actual number. 

A crucial aspect of bubbles is that they can be used as calorimeters. This means that bubbles give a means to derive the energy being injected into the medium by the AGN. This energy can then be compared to the energy needed to offset cooling of the hot ICM. The energy 
associated with
a bubble can be determined by calculating its enthalpy, which is the sum of the work needed to displace the gas and the thermal energy within the bubble \citep{1973Natur.244...80G,2000A&A...356..788C}.

\begin{eqnarray}
E_{\rm cav}= E_{\rm thermal} + PV =\frac{\gamma PV}{\gamma - 1}
\label{eq:ecav}
\end{eqnarray}

Here, $P$ is the pressure surrounding the bubble and is usually taken as the azimuthally averaged pressure as derived by X-ray observations at the location of the cavity \citep{2006ApJ...652..216R}. $V$ is the volume of the cavity, and a spherical or prolate shape is assumed. For a prolate shape, $V=4\pi \times R_{\rm w}^{2}R_{\rm l}/3$, where $R_{\rm l}$ is the semi-major axis along the direction of the jet, and $R_{\rm w}$ is the semi-major axis perpendicular to the direction of the jet. We assume that bubbles are filled with an ideal gas, where $\gamma$ is the ratio of specific heats (4/3 for relativistic gas and 5/3 for non-relativistic gas). Essentially, it is thought that bubbles can provide energies ranging from $2.5PV$ ($\gamma=5/3$) to 4PV ($\gamma=4/3$). These energies can reach $10^{62}$ erg in some clusters. 

The power injected into the medium is then determined by dividing the energy with the bubble’s age. This age is given as the buoyant rise time, the refill time, or the sound crossing time. See respectively the equations below \citep{2000A&A...356..788C,2001ApJ...562L.149M,2007ARA&A..45..117M}.

Here, R is the distance from the radio point source to the middle of the cavity (projected), S is the cross-sectional area of the bubble ($S = \pi R_{\rm w}^2$), $C_{\rm D}=0.75$ is the drag coefficient \citep{2001ApJ...554..261C,2018MNRAS.478.4785Z}, $g$ is the local gravitational acceleration ($g = GM(<R)/R^2$), $r$ is the bubble radius ($r = \sqrt{R_{\rm l} R_{\rm w}}$ for a prolate bubble) and $c_{\rm s}$ is the sound crossing time ($c_{\rm s} = \sqrt{\gamma kT/(\mu m_{\rm H})}$, where $kT$ is the plasma temperature at the radius of the bubble, $\gamma = 5/3$ and $\mu = 0.61$).

\begin{eqnarray}
t_{\rm buoyant} = \frac{R}{v_{\rm terminal}} = R\sqrt{\frac{S C_{\rm D}}{2gV}}\approx 0.5 \frac{R}{r}\sqrt{\frac{r}{g}}
\label{eq:tbuoy}
\end{eqnarray}
\begin{eqnarray}
t_{\rm refill}= 2\sqrt{\frac{r}{g}}
\label{eq:tref}
\end{eqnarray}
\begin{eqnarray}
t_{\rm cross}= \frac{R}{c_{\rm s}}
\label{eq:tcross}
\end{eqnarray}

The buoyant rise time is the time it takes a bubble to reach  its current position with the terminal buoyant velocity ($v_{\rm terminal}$), which depends on medium drag forces. This is a valid estimate of the X-ray cavities’ age for the ones that have clearly detached from their AGN and have risen. The refill time is the time it takes a bubble to rise buoyantly through a distance equal to its own diameter. Finally, $t_{\rm cross}$ shows either a pressure equilibration time or the bubble expansion at trans-sonic velocities. On shorter time scales the role of buoyancy is small.


\begin{figure}[!ht]
\includegraphics[trim=0cm 0cm 0cm 0cm,width=1.0\columnwidth]{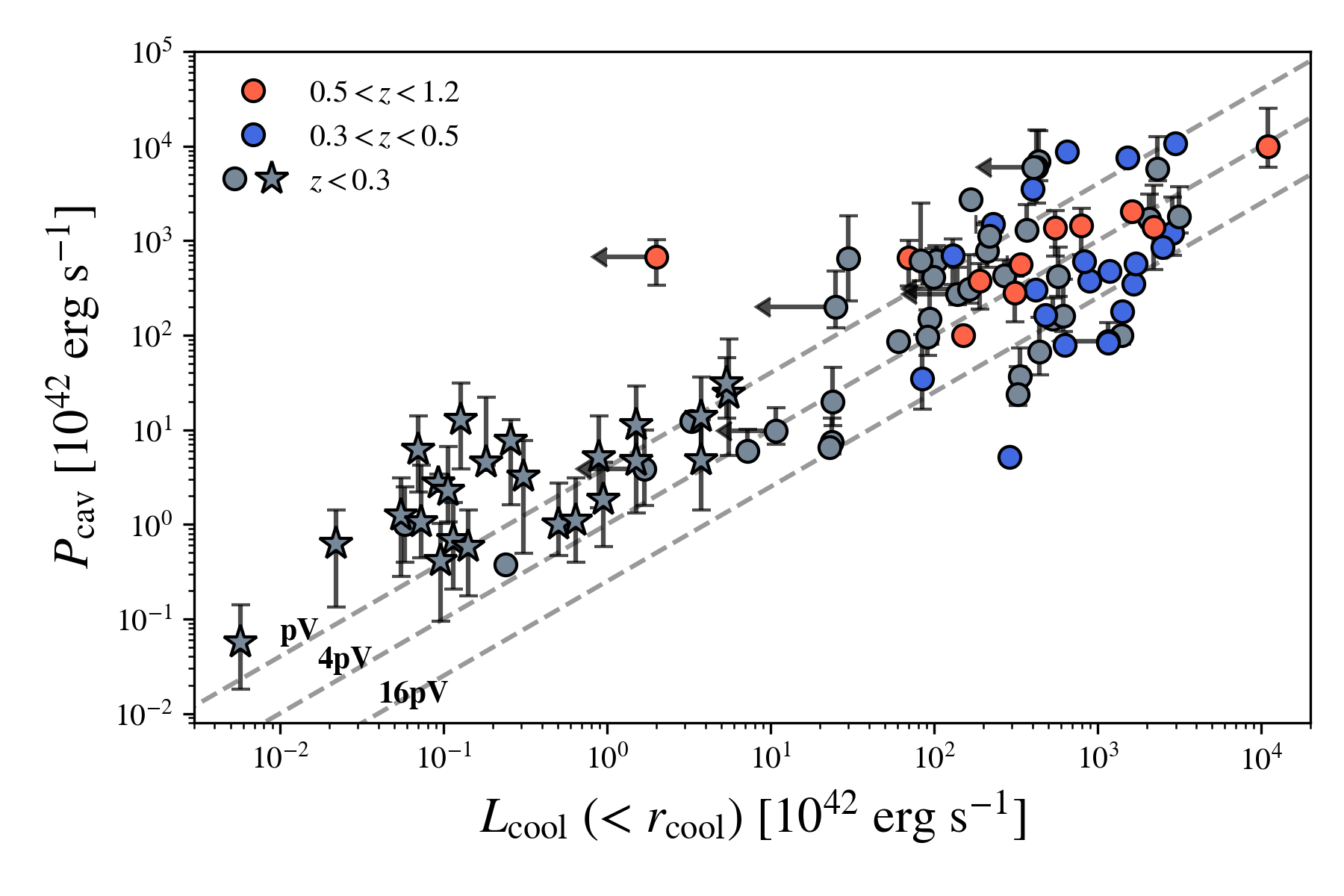}
\caption{Total mechanical power ($P_{\rm cav}$) of X-ray cavities as a function of the cooling luminosity ($L_{\rm cool}$). The latter is defined as the X-ray luminosity contained within a radius where the cooling time is equal to 7.7 Gyrs. Data for more isolated systems and groups of galaxies are shown with grey stars \citep{2009AIPC.1201..198N}. They are all located at $z<0.3$. The remaining data are for clusters of galaxies and are colour coded depending on their redshifts. The data points are from \citet{2006ApJ...652..216R}, \citet{2008MNRAS.385..757D}, \citet{2017MNRAS.466.2054V,2019MNRAS.485.1981V}, \citet{2021SerAJ.202...17K}, \citet{2013Ap&SS.345..183P,2019ApJ...870...62P,2021MNRAS.504.1644P},\citet{2019MNRAS.484.4113K},\citet{2020MNRAS.492.3156L},\citet{2015Ap&SS.359...61S},\citet{2012MNRAS.421.1360H},\citet{2015ApJ...805...35H},\citet{2020AJ....160..103P} and \citet{2019ApJ...885...63M}. Note that the time scale used to determine $P_{\rm cav}$ mostly consisted of the time scale as defined by Eq. \ref{eq:tbuoy} and Eq. \ref{eq:tcross}, but that the value was not always reported in the literature. Dashed black lines denote $P_{\rm cav}=L_{\rm cool}$ for $E_{\rm cav} =$ PV, 4PV and 16PV . Credit: M. Prasow-Émond.}
\label{fig:PcavLCool}
\end{figure}

One of the most important results from studying AGN feedback is that AGN seem to inject on average the required amount of energy needed to offset cooling in the ICM. This is shown in Figure \ref{fig:PcavLCool}. Here, the cavity power of the AGN is plotted against the cluster X-ray luminosity within the cooling radius. This figure shows that most AGN have X-ray cavities that can energetically offset cooling (some even have a surplus of energy). In other words, according to this figure and the fact that many clusters have short cooling times near their center, there is a fine balance between heating and cooling. If there was no additional heating, more cold gas and soft X-rays would be seen. If there was too much additional heating, then the central entropy would increase up to a point where the central cooling times would approach typical cluster ages.

On the scales of the galaxies, the primary impact is that AGN feedback prevents stars from forming by heating the surrounding medium. AGN feedback can also in principle halt star formation by expelling gas from the galaxy, but most of the outflows seen - for radio mode feedback - are below the escape velocity of the galaxy (see Section \ref{obsoutflows}). Therefore, radio mode feedback might be responsible of ``stirring'' things around in a galaxy, but it does not expel gas into the intracluster gas. If such processes happen on long time scales, then the impact on galaxies is colossal: according to cosmological simulations (without feedback), massive galaxies (weighing above $M_\star\sim10^{11}M_\odot$) would be tens to hundreds times more numerous than observed without AGN feedback occurring \citep{2018MNRAS.475..648P}.

\subsubsection{Heating by shocks, mixing, turbulence and/or sound waves}\label{obsheat} 

Let us consider the example mentioned earlier, about the balloon being inflated in a room. By inflating a balloon in a room, as one can experiment, this process does not automatically heat the room, so how can black holes - by inflating bubbles via their radio jets - heat the surrounding hot gas and prevent it from cooling? 

Initially, it was not clear how this process occurred, but more recently, we have started to have a better understanding.

One might first argue that the energy must be transferred primarily via shocks as the supersonic radio jet propagates and pushes away the hot gas. Yet, it appears that radio jets do not generate strong shock fronts in galaxy clusters - instead they seem to drive very weak shock fronts \citep[e.g. ][]{2005ApJ...628..629N,2007ApJ...665.1057F,2009A&A...495..721S}.


These shock fronts are visible at X-ray wavelengths and typically, have low Mach numbers (Mach $= 1-2$), making them very difficult to see in X-ray observations (see also Chapter on the dynamics of the X-ray emitting plasma in clusters). They are also usually located near the bubbles \citep[see for example the cases of NGC 5813 and Perseus; ][]{2008MNRAS.386..278G,2015ApJ...805..112R}. Even with low Mach numbers, these shocks can carry a substantial amount of energy (typically $10^{44-45}$ erg s$^{-1}$), often enough to suppress cooling of the hot X-ray gas in the very core of the cluster.

On larger scales (i.e. on the scale of the entire cool core), the details are less clear. Some possibilities include heating by sounds waves \citep[e.g. ][]{2006MNRAS.366..417F,2017MNRAS.464L...1F}, cosmic ray streaming \citep[e.g. ][]{2008MNRAS.384..251G}, mixing \citep[see e.g. ][]{2020ApJ...896..104H} or turbulence \citep[e.g. ][]{2014Natur.515...85Z}. The latter hypothesis advanced thanks to a large effort in the community which led to the seminal work that measured amplitude fluctuations in the \textit{Chandra} X-ray images of nearby galaxy clusters \citep{2014Natur.515...85Z,2015MNRAS.450.4184Z,2019SSRv..215...24S}. This method assumes that the amplitude fluctuations scale linearly with the velocity spectrum, and hence can allow for a measurement of turbulence. \citet{2018ApJ...865...53Z} found in particular that the turbulence measured in the inner $30-40$ kpc from X-ray surface brightness and gas density fluctuations is in agreement with the velocity broadening measurements made by the \textit{Hitomi} satellite for the Perseus cluster \citep[164 km s$^{-1}$$\pm10$ km $s^{-1}$][]{2016Natur.535..117H}, and that this amount of turbulence can be sufficient to offset cooling of the hot X-ray gas. More recently, it has been hypothesized that the central AGN in galaxy clusters may not be the only source of heating of the cool core. Indeed, turbulence measurements in the outer parts of the cluster (i.e. beyond $>60$ kpc) can be accounted for by cluster formation processes, which include sloshing motions caused by minor mergers. Such turbulent motions could theoretically transfer from the outskirts down to the core of the cluster, and offset the cooling of the cool core altogether. 

More insight into the key mechanism of heating will be made possible with the advent of upcoming X-ray satellites, noticeably the X-Ray Imaging and Spectroscopy Mission (\textit{XRISM}) scheduled for 2022, as well as the \textit{Athena} (Advanced Telescope for High ENergy Astrophysics) observatory scheduled for the next decade and the mission concept \textit{Lynx} X-ray observatory. 

\subsubsection{Radio jets and massive molecular outflows}\label{obsoutflows}

Over the past two decades, a substantial effort has been devoted to understanding AGN-driven outflows, especially those associated with quasars in which the black hole is accreting at high rates. 
One of the reasons astronomers started studying such AGN, is that large outflows were often seen to be associated with them. These outflows appear to be highly multi-phase and include neutral atomic gas outflows reaching velocities up to 1500 km s$^{-1}$ \citep[e.g. ][]{2005A&A...444L...9M}, mildly ionized warm absorbers with velocities of a few hundred km s$^{-1}$ \citep[e.g. ][]{2005A&A...432...15P}, to ionized outflows reaching several thousand km s$^{-1}$ as seen from optical/UV broad absorption line systems or [OIII] broad emission lines and even highly ionized outflows reaching 0.2-0.4c as seen from Fe K-shell absorption \citep[e.g. ][]{2011ApJ...742...44T}. Some outflows appear to be concentrated around the AGN, but others can reach kpc scales and hence may have a profound impact across the entire galaxy. Quasars also appear to be able to propel molecular gas at high speeds (several thousand km s$^{-1}$) on kpc scales, as seen in some highly obscured AGN in dusty star-forming galaxies \citep[see ][]{2013ApJ...776...27V}.

In the case of galaxy clusters, the central AGN is often radiatively inefficient and in radio mode feedback \citep[e.g. ][]{2011MNRAS.413..313H}. Most of the feedback originating from such a black hole therefore occurs from the interaction between the radio jet and the surrounding gas. The Phoenix cluster of galaxies is however one example hosting a central quasar \citep[][ and references therein]{2012Natur.488..349M}.  As mentioned earlier, this cluster harbors a central AGN which is accreting at high rates (in quasar-mode), yet is also driving powerful radio jets \citep[e.g. ][]{2015ApJ...805...35H}. The central galaxy also shows evidence of a highly ionized ([OIII], He II) outflow with $<\sigma_{\rm v}>\sim300$ km s$^{-1}$ located roughly 20 kpc North of the central galaxy. The outflow however appears to coincide with the X-ray cavities being created by the radio jets - the outflow may therefore originate from the radio jets that heat the surrounding medium via shocks. 

For some time, it was thought that radio jets simply drill through a galaxy and that most of the energy must be deposited outside of the host. However, it has now become clear that radio jets play a much more fundamental role outside and throughout the entire galaxy. 

Initial studies showed that radio jets may entrain metals with them, often out to dozens of kpc as in the case of MS 0735.6+7421, which shows metal-enriched gas along the 200 kpc jets \citep[see ][]{2011ApJ...731L..23K}. Similarly, there is also now clear evidence that radio jets can also entrain atomic and molecular gas leading to massive outflows of several hundreds of km s$^{-1}$ \citep[e.g. ][]{2005A&A...444L...9M,2011ApJ...729L..27R}.

Recent studies have shown that such gas can be uplifted by the buoyantly rising X-ray cavities \citep[e.g. ][and references therein]{2014ApJ...785...44M,2014ApJ...784...78R}. This is in part due to the advent of ALMA. With its unprecedented sensitivity and resolution, ALMA has revealed the existence of cold gas in thin-like filaments that often surround X-ray cavities or trail behind them. These filaments can reach dozens of kpc in size (see for example the case of the Perseus cluster with filaments reaching scales of 100 kpc). The masses at play are often on the order of a few to tens of billions of solar masses, which rivals even the most gas-rich galaxies at high-redshift. The connection to radio jets is somewhat unexpected given that molecular gas was initially thought to act like a barrier that would stomp waves or redirect the radio jets. Instead, radio jets seem to play a fundamental role in ``stirring up'' the galaxy - and this may not only be the case of the central cluster galaxies, but may also be a more wide-spread phenomenon in all galaxies. 

A fantastic example of this process is shown in the case of Abell 2597 ($z=0.0821$). Abell 2597 is a nearby cool core cluster that harbors a typical pair of X-ray cavities (approximately 30 kpc in size), as well as the usual multi-phase filamentary nebula (also approximately 30 kpc in size) seen in the cores of many cool core clusters. These nebulae are extremely thin \citep[e.g. ][]{2008Natur.454..968F}, most likely made of cold molecular clouds that are surrounded by warmer ionized gas, and are often remarkably co-spatial with the X-ray cavities. 

The extensive multi-wavelength coverage of Abell 2597, noticeably ALMA and MUSE observations, paint a clear picture of how AGN feedback operates in the cores of clusters \citep[see in particular][]{2018ApJ...865...13T}. These data show that the warm ionized and cold molecular components are highly co-spatial and co-moving. They appear to not be in dynamical equilibrium and instead, show evidence of inflow towards the central black hole, outflow along the radio jets and uplift by the buoyantly rising X-ray cavities (that are created by the radio jets). This is entirely consistent with a galaxy-spanning ``fountain'' (a term that the authors use) that illustrates a remarkably tight self-regulated feedback loop between the central AGN and the surrounding hot X-ray halo. The cold gas rains down onto the central black hole, which powers radio jets that inflate X-ray cavities. These cavities then drive turbulence, shocks and/or sound waves, in addition to uplifting low entropy multi-phase gas to larger radii. The latter then triggers additional cooling of the clouds. The velocity of such clouds is well below the escape velocity, which implies that the clouds will then rain back down onto the central black hole, re-triggering a new feedback loop. 

One might argue that the AGN feedback loop occurring in Abell 2597 must be unique to galaxy clusters and that such loops are only established in these prime circumstances, but what is emerging from the literature, is that the AGN feedback loop we are witnessing may be a truly fundamental process of galaxy evolution in essentially all galaxies. Therefore, as long as a galaxy harbors a black hole (which most seem to do) and this galaxy contains intergalactic gas (which most seem to), then this tightly self-regulated loop may be one of the main fundamental processes that govern star formation and galaxy growth over time.

\subsubsection{The evolution of AGN feedback in groups and clusters of galaxies}\label{obsevo}

The question which we now address is whether AGN feedback still plays a fundamental role in high-redshift (z$\gtrsim0.3$) clusters of galaxies? 

Observationally, there are two ways of addressing this question. The first relies on finding direct evidence of AGN feedback occurring in more distant groups and clusters. Some studies have succeeded in doing this by using deep X-ray observations of distant clusters of galaxies in search of X-ray cavities or deep radio observations in search of radio jets originating from the central AGN \citep[see e.g. ][]{2015ApJ...805...35H,2012MNRAS.421.1360H,2019ApJ...887L..17C}. This has been possible only recently thanks to the tremendous efforts that have been deployed by the community in search of distant clusters of galaxies. Such samples include those identified via the SZ effect with the South Pole Telescope \citep[][]{2011PASP..123..568C}) and the Atacama Cosmology Telescope \citep[ACT; ][]{2011ApJS..194...41S}, both of which have discovered hundreds of $z>1$ clusters. The \textit{Spitzer} Adaptation of the Red-Sequence Cluster Survey (SpARCS) and Stellar Bump Sequence survey have also discovered over 500 $z > 0.6$ clusters in the \textit{Spitzer} Wide-area Infrared Extragalactic (SWIRE) fields \citep[e.g. ][]{2009ApJ...701.1839M,2009ApJ...698.1943W}. Combined with  large targeted X-ray follow-up programs, it has been possible to search for X-ray cavities systematically in large high-redshift samples of clusters.

Overall, these studies find evidence of $\sim$10-30 kpc-scale X-ray cavities in galaxy clusters located between $z\sim0.3$ to $z\sim1$. Considering that the first galaxy clusters formed at about $z\sim2$ (i.e. when the Universe was a couple billion years old), this implies that powerful AGN feedback - in radio mode - may have been operating in clusters of galaxies for most of their lifetime. There are however caveats. Indeed, these studies are limited by the quantity of available data, since only a handful of clusters have deep enough X-ray observations to find X-ray cavities at high redshift. Although $\textit{Chandra}'s$ spatial resolution can resolve $\sim$20 kpc-scale X-ray cavities out to $z\sim2$, the smaller ones will be missed. These studies are also biased towards identifying X-ray cavities in cool-core clusters, given that the X-ray counts will be highly concentrated in the core of these clusters. This makes identifying X-ray cavities in cool core clusters easier, given that cavities are usually located in the cores. 

Indirectly, by studying the evolution of ICM properties, there are now multiple studies that support the presence of AGN feedback processes in the cores of high-redshift clusters of galaxies. For example, by using a large sample of clusters of galaxies discovered by the South Pole Telescope, \citet{2013ApJ...774...23M,2014ApJ...794...67M} found that the electron number density of the ICM appears to be evolving in a self-similar manner, except for the profiles in the cores of clusters which appear to deviate significantly from self-similarity. 

In fact, further studies at even higher-redshift ($z>1.2$) confirm earlier results of no evolution in the cluster cores, implying that AGN feedback processes must already be in place at these redshifts in the cluster cores \citep{2017ApJ...843...28M}. A similar study, this time focusing on combining \textit{Chandra}’s spatial resolution and \textit{XMM-Newton}’s large field of view and effective area, show the same results for the cores of galaxy clusters \citep{2021ApJ...910...14G}. Overall, these studies seem to indicate that the fine balance between AGN feedback and ICM cooling in the cores of galaxy clusters has not evolved since z$\sim1-2$.

On the other hand, a recent study seems to suggest that AGN feedback may not be operating for all cool core galaxy clusters at high-redshift. Indeed, the $\sim10^{14}M_\odot$ galaxy cluster SpARCS104922.6+564032.5 located at $z=1.709$ \citep{2015ApJ...814...96W} appears to host an extreme starburst fuelling $\sim900M_\odot$ of star formation in the core. Based on \textit{Chandra} observations, \citet{2020ApJ...898L..50H} showed that this burst of star formation appears to be fuelled by runaway gas cooling for the ICM (i.e. a true cooling flow) occurring in the absence of AGN feedback. In less than 100 million years, such runaway cooling can form the same amount of stars as in the Milky Way, implying that intracluster stars are not only produced by tidal stripping and the disruption of cluster galaxies, but can also be produced by runaway cooling of the ICM at early times. This study would therefore be the first example of a true cooling flow and may signify a new era of AGN feedback processes occurring at cosmic noon (z$\sim2$). More detailed studies of clusters and protoclusters at z$>2$ are needed to explore this in more detail.

\section{Models of AGN feedback}\label{sec:models}
To date, many theoretical models of AGN feedback have been considered, motivated by rich observational evidence described above. In particular, the following properties of the cores of relaxed groups and clusters of galaxies were the main drivers of these models: 
\begin{itemize}
\item Short cooling time of the gas in the core $t_{\rm cool}\lesssim 10^9\,{\rm yr}\ll t_{\rm H}$
\item No evidence for high star formation rates that might be expected for the mass cooling rate derived from the observed radiative losses $SFR\ll \dot{M}\sim\frac{L_{\rm X}}{kT}m_{\rm p}\sim 1000\,M_\odot\,{\rm yr^{-1}}$ 
\item Lack of emission lines characteristic for gas cooling well below $\sim$1 keV.    
\item Presence of a low-luminosity AGN (radio source) in the central elliptical galaxy
\item Evidence for radio-bright bubbles (= X-ray cavities) in the core with inferred power comparable to the cooling losses $\sim L_{\rm X}$.
\end{itemize}
Given that the gas radiative losses are directly observed in the X-ray band, and yet there are no signs of gas cooling to low temperatures, the only viable scenario\footnote{Hiding the cooling/cold gas in some ``invisible'' form could be another option, but no compelling evidence for this has been found so far.} is to assume that some source of energy is compensating the losses. Since radio sources (central AGN) are invariably found in the cores of the relaxed clusters and the amount of energy they release in a form of jets/outflows is large enough, it is plausible that this is the source of energy one is looking for. 
This is the foundation of the AGN feedback model, directly supported by observations  and the energetic estimates. Moreover, AGN-induced bubbles in groups and clusters with orders of magnitude different luminosities, have very similar appearances (apart from their physical size) and the AGN power apparently scales with the gas cooling losses. This implies that a system consisting of a cooling atmosphere and an AGN are able to communicate with each other and establish an approximate balance between cooling and heating. For this to work, general arguments about available power/energy are not sufficient. The gist of the radio mode AGN feedback model is that the following loop is operating in cluster cores:

{\bf gas heating is insufficient to prevent cooling $\rightarrow$ higher SMBH energy release rate $\rightarrow$  more gas heating  }

This loop can be viewed as an example of a system with negative feedback that may naturally live in a state close to an equilibrium or oscillate around it. However, there is no guarantee that in real clusters this loop works, and a careful assessment of each step is needed to prove the concept. In particular, any reasonably complete AGN feedback model should answer the following questions:
\begin{itemize}
\item Q1: What is the most relevant black hole accretion mode?
\item Q2: How does the gas cooling on large scale affect the accretion rate?
\item Q3: In what form is the energy released by the black hole?
\item Q4: How is this energy dissipated into gas heating?
\end{itemize}

Remarkably, a variety of answers to these questions (and their combinations) has been suggested and claimed to be successful. The reason for such a diversity stems from our inability to model the clusters and black holes from the first principles. While this at first glance seems problematic, it also shows that the final answer (effective negative feedback) is robust to detailed assumptions. This is also seen in numerical simulations (see Section~\ref{sec:sim1}). 

\subsection{Feeding the AGN}\label{accretionEugene}
Broadly, two accretion modes are considered: the so-called ``hot'' and ``cold'' modes. In the former case, hot gas from an almost hydrostatic atmosphere is captured by a black hole sphere of influence and flows on the BH. In contrast, in the ``cold'' mode, the gas cools much below the virial temperature of the system before it is captured by the BH. In both cases, excessive cooling of the gas might lead to the increased accretion rate and, plausibly, to the increased amount of energy available for gas heating. In response to the gas accretion, the BH can release energy as radiation or in a mechanical form (including jets and winds), the partition between different forms being a function of the accretion rate and mode, and, possibly, of other parameters. In particular, the radiation is expected to be subdominant at low accretion rates, when the mechanical energy (predominantly jets) is instead more important. At high accretion rates, a radiatively efficient accretion flow is expected, although the mechanical energy release (e.g., powerful winds) might be present too.  The mechanical energy has an apparent advantage of a more efficient coupling with the gas than radiation and many models of the AGN feedback in clusters rely on this form of the BH energy release, although different routes of energy dissipation are possible.         

Conceptually, one of the  simplest sets of answers (not necessarily correct ones) to the above questions is:
\begin{enumerate}
\item A1: The accretion rate is set by the Bondi model (spherically symmetric accretion by a point mass embedded into a uniform medium), which is the simplest version of the ``hot'' accretion mode. 
\item A2: Cooling of the gas leads to the decrease of the gas entropy, which in the Bondi model directly controls the accretion rate.   
\item A3: SMBH releases energy quasi-continuously in a mechanical form and inflates bubbles filled with very hot plasma (e.g. cosmic rays).  
\item A4: The bubbles rise buoyantly in the cluster atmospheres and transfer their energy to the gas. This energy is eventually dissipated into heat. The details of this process are complicated, but the energy conservation law ensures that this will happen.  
\end{enumerate}
We will go below through this simplified picture before mentioning the plethora of other options.

The Bondi accretion is a classic problem of accretion onto a point mass in a homogeneous medium \citep{1952MNRAS.112..195B}.  Although it is highly idealized and, in particular, ignores the inevitable angular momentum of the gas, it is a very useful model to consider. The Bondi rate can be estimated as a transonic flow of the gas with density $\rho$ through the sphere with the Bondi radius $r_{\rm B}$  
\begin{eqnarray}
\dot{M}\approx \pi r_{\rm B}^2 c_{\rm s} \rho \approx \pi (GM_{\rm BH})^2 c_{\rm s}^{-3}n \mu m_{\rm p} \propto M_{\rm BH}^2\frac{n}{T^{3/2}}\propto M_{\rm BH}^2 s^{-3/2},
\end{eqnarray}
where $s=\frac{T}{n^{3/2}}$ is the gas entropy index.  Here $r_{\rm B}$ is the Bondi radius 
\begin{eqnarray}
r_{\rm B}=\frac{2GM_{\rm BH}}{c_{\rm s}^2}\approx 0.1 \left( \frac{M_{\rm BH}}{10^{10}\,M_\odot} \right)\left( \frac{kT}{3\,{\rm keV}} \right)^{-1}\,{\rm kpc},
\label{eq:r_B}
\end{eqnarray}
where $c_{\rm s}$ is the sound speed of the gas, which for the adiabatic equation of state is equal to  $\left (\gamma\frac{kT}{\mu m_{\rm p}} \right)^{1/2}$.  At the Bondi radius the characteristic thermal energy of a gas particle $\sim kT$ equals to its potential energy due to BH gravity $\sim m_{\rm p}\frac{GM_{\rm BH}}{r_{\rm B}}$.
Therefore, in the frame of the Bondi accretion model, the mass accretion rate directly depends on the entropy of the gas $s$. As discussed in the next Section, due to  simplicity and physical transparency of the Bondi recipe, it is often used in numerical simulations of the AGN feedback.

The models based on the Bondi accretion have another attractive feature. Namely, they can immediately illustrate the connection between the gas cooling on large scales and the SMBH accretion rate. The Schwarzschild radius of a $10^{10}\,{\rm M_\odot}$ black hole is $r_{\rm s}=2GM/c^2\sim 3\,10^{15}\,{\rm cm}$ or, equivalently, $\sim 10^{-6}\,{\rm kpc}$, i.e. much smaller than any spatial scales that characterize the size of the region  where the majority of the cooling losses occur ($r_{\rm cool}\sim 100-200$~kpc in massive clusters). The Bondi radius (see eq.~\ref{eq:r_B}) is also very small compared to $r_{\rm cool}$. For conditions typical for groups and clusters, within the Bondi radius, the BH gravity dominates over the contributions of other components, such as dark matter, stars or gas, or is at least comparable to them. However, on larger scales the black hole contribution to the mass/potential is small. All in all, the BH can control the gas motions inside the Bondi radius, but its gravity has little impact on the gas farther out, where most of radiation losses occur in real clusters. However, for the radio mode feedback to work efficiently (= to achieve an approximate balance between cooling and heating), the cooling gas and the BH should somehow communicate with each other. Given the large scale separation $r_{\rm cool} \gg r_{\rm B} \gg r_{\rm s}$ this, at first glance, seems problematic. However, the ``central'' black hole in a relaxed cluster is located at the center of a giant elliptical galaxy (typically much more massive than other galaxies in the cluster), which itself is located at the center of the cluster potential well. In an approximately hydrostatic atmosphere, this guarantees that the  lowest-entropy gas in the halo will eventually end up at the bottom of the potential well\footnote{Note that this requires the cooling time of the gas to be substantially longer than the sound crossing time, i.e., $t_{\rm cool}\gg r/c_{\rm s}$. In the opposite case, the gas can cool ``in place'' before it settles into hydrostatic configuration. We return to this question below.}, where the BH is located. Therefore the entropy index $s$ that enters the expression for the Bondi rate represents the lowest entropy present in the system. In other words, if cooling dominates over heating and the entropy of the gas goes down, this low entropy gas will appear in the vicinity of the black hole and will drive the accretion rate up. This way, the BH becomes aware of the gas properties on much larger scales. Basically, the BH ``monitors'' the presence of the low entropy gas in the systems and is trying to counteract further entropy decrease - a variant of a negative feedback loop.
Thus, the Bondi-type accretion provides very elucidative example of a model with a direct link of the SMBH accretion rate to the gas cooling-heating balance on vastly larger spatial scales. In this model, the system can evolve into a balanced state, when cooling and heating are approximately equal. Moreover, this balanced state can be stable, although the question of global stability requires detailed modeling. 

The question of the Bondi formula applicability is an actively debated issue, which is not fully resolved yet. In section \ref{BondiJulie}, we address this from an observational perspective, which illustrates how Bondi accretion may not work for all objects. There are open questions on the theory side too. Firstly, for the flow that conserves the Bernoulli parameter, the amount of matter reaching the black hole can be much smaller (due to outflows) than the amount of matter entering the Bondi radius \citep{1999MNRAS.303L...1B}. Secondly, non-zero angular momentum can prevent gas from falling onto BH radially, forming a rotationally-supported structure instead. On the other hand, under reasonable assumptions (high effective viscosity and relatively small angular momentum) one can construct a flow that starts well outside the Bondi radius and leads to SMBH accretion rates not much different from those based on the Bondi model \citep{2011MNRAS.415.3721N}.  

An alternative to the ``hot'' accretion is the so-called ``cold'' accretion mode \citep{2005ApJ...632..821P,2010MNRAS.408..961P},  when the gas first cools and forms denser clumps, which lose their angular momentum and eventually feed the BH. In this case, the accretion rate onto the BH is no longer determined determined by the hot gas entropy and can be much higher than the Bondi value. It is also plausible that the accretion rate in this mode is strongly variable as demonstrated by numerical simulations (see Section~\ref{sec:sim1}). There is no final consensus on which of the two accretion modes (hot or cold) provides the bulk of the needed power to the ICM. Conceptually, the hot mode is simpler, while the cold mode has its own interesting and attractive features. Since the black hole ``does not care'' for the gas but simply responds to the level of feeding, it is possible to imagine that different modes are operating in different objects (or at different times in the same object). There is, however, a consensus that the excessive cooling of the gas leads to the increase of the accretion rate onto SMBH in both cases, allowing for a negative feedback loop.

Both types of AGN feeding (and their variants) are the subjects of ongoing studies and it is plausible that different forms of feeding prevail in different classes of objects.

\subsection{Energy release by supermassive black holes}
The next important question is the relation between the accretion rate and the energy released by the black hole. The energetics arguments (\S\ref{sec:intro}) require highly efficient conversion of the released energy into gas heating and provide an estimate of the time-averaged accretion rate. If the conversion efficiency is indeed very high, the time-averaged power released by the AGN can be lower than the Eddington luminosity for a reasonably massive black hole (see Eq.~\ref{eq:redd}). 

There is a large body of observational, theoretical, and numerical studies focused on the physics of accretion and associated energy release \citep{2002apa..book.....F,2014ARA&A..52..529Y}. In the most simple form, many basic characteristics of the BH energy release can be linked to its accretion rate scaled by the Eddington rate:
\begin{eqnarray}
\dot{m}=\frac{\dot{M}}{\dot{M}_{\rm Edd}}=\frac{\dot{M}\sigma_{\rm T}c\epsilon_0}{4\pi G M_{\rm BH}m_{\rm p}},
\label{eq:mdedd}
\end{eqnarray}
which assumes that the gas opacity is set by the Thomson scattering of photons by electrons, $\sigma_{\rm T}$ is the Thomson cross section and $\epsilon_0$ is the radiative efficiency of the accretion flow. While this approach ignores many important factors like the spin of the black hole,  angular momentum, and magnetic field of the accreted gas, it, nevertheless, captures major differences between the very low and high accretion rates. It also helps to consider stellar mass black holes together with their supermassive analogs. Observations of stellar mass black holes in binary systems bring in important information since on time scales of days, months, and years their accretion rates change by several orders of magnitude. This allows  probing very different accretion regimes in the same object \citep{2004MNRAS.355.1105F}. One of the most salient differences between the low and high accretion rates is their radiative efficiency. At high accretion rates, an optically thick, but geometrically thin accretion disk (often called Shakura-Sunyaev disk \citep{1973A&A....24..337S}) is present with high radiative efficiency. For non-rotating black holes, its radiative efficiency can correspond to the canonical value of the binding energy of matter at the innermost stable orbit $L/(\dot{M}c^2)\approx 0.06$ (and up to 0.42 for maximally rotating black hole). For stellar-mass binaries, this is the so-called ``high state''; supermassive black holes in this regime appear as bright sources known in astronomy as Quasi-Stellar Objects (QSO).  

In the opposite limit of a much lower accretion rate [less than a few per cent of the Eddington value, i.e. $\dot{m}<\dot{m}_{\rm crit}\sim O(10^{-2})$], the flow is geometrically thick but optically thin, at least in the inner region. In binaries, this is the so-called ``low state'', while for SMBH it corresponds to  the class of low luminosity AGN. In this state, the low density of the flow prevents its cooling and formation of a thin disk. The bulk of the gas energy is not radiated hence they are often called Radiatively Inefficient Accretion Flows (RIAF; other names reflecting the roles of advection, outflows or convection specific for different models are also used)   \citep{1977ApJ...214..840I,1994ApJ...428L..13N,1995ApJ...438L..37A,1999MNRAS.303L...1B}. To the first approximation, the radiative output in this regime is proportional to $\varepsilon_0\dot{m}\times\left(\dot{m}/\dot{m}_{\rm crit} \right)$ and can be much lower than the radiatively efficient QSO regime, where it is $\sim\varepsilon_0 \dot{m}$.  

\begin{figure} 
\includegraphics[trim=4cm 2cm 0cm 0cm,width=0.5\columnwidth]{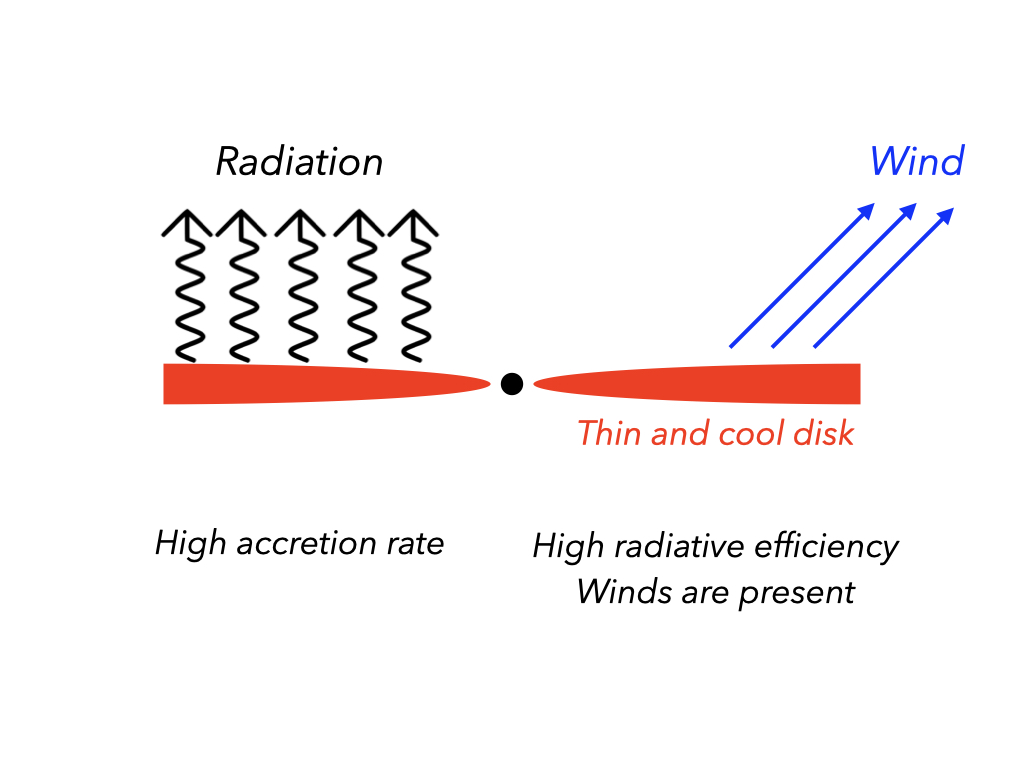}
\includegraphics[trim=0cm 0cm 0cm 0cm,width=0.5\columnwidth]{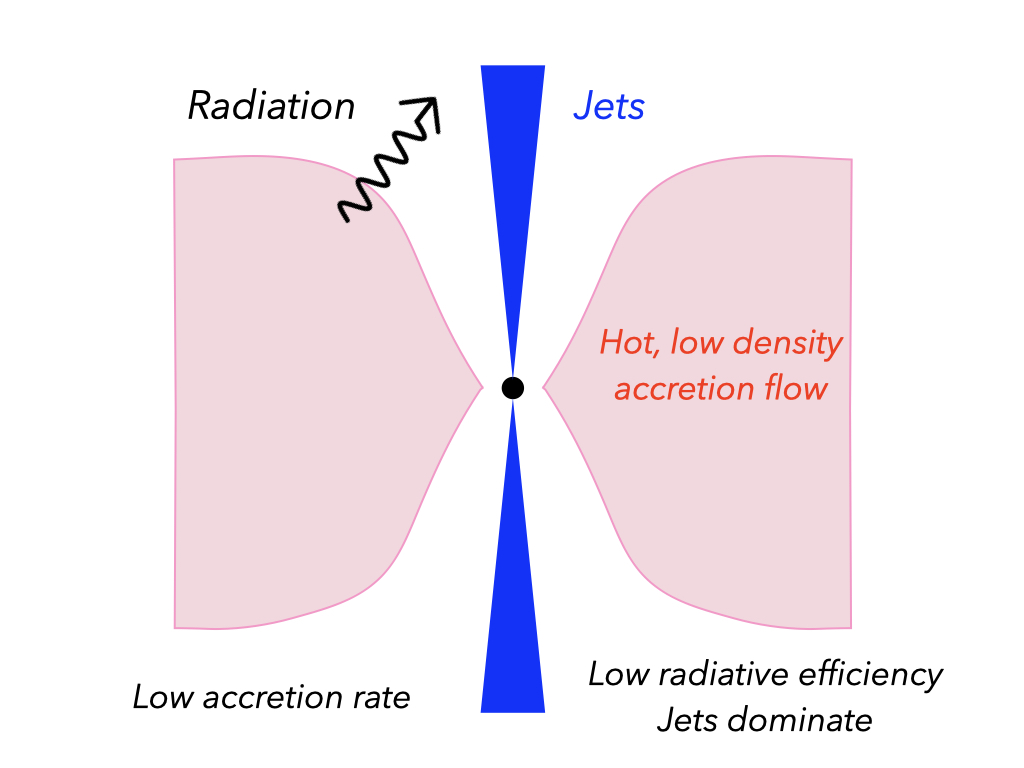}
\caption{Standard cartoon of inner regions of the accretion flow configurations at high and low accretion rates (left and right, respectively). At high accretion rates, an optically thick and geometrically thin accretion disk is present. The flow is radiatively efficient and can launch  winds/outflows. In contrast, at low accretion rates (below a few \% of the Eddington rate), the flow is optically thin and geometrically thick. The radiative efficiency is small and more energy is channeled into jets. The latter configuration is the most relevant for SMBHs in the present day clusters.}
\label{fig:acc_hl}
\end{figure}

Apart from radiation, a fraction of the SMBH energy release can be in a mechanical form, namely carried by particles rather than photons. 
At high accretion rates (radiatively efficient QSO-like state), fast wide-angle winds are ubiquitously seen in observations and also produced in simulations \citep{2000ApJ...543..686P,2015ARA&A..53..115K}. Powerful AGN radiation is likely playing a role in the acceleration of these winds \citep{1983ApJ...271...70B}. The estimates of the winds power are uncertain, but could be a sizable fraction of the radiative output or even comparable to it. In addition to winds, some 10\% of QSO are radio loud and possess well collimated radio jets. 

In the opposite limit of low accretion rates (radiatively inefficient modes, \citealt{2014ARA&A..52..529Y}), both simulations \citep{2011MNRAS.418L..79T,2012MNRAS.423.3083M,2016MNRAS.461L..46T} and observations \citep{2004MNRAS.355.1105F,2003MNRAS.345.1057M} show that collimated relativistic jets are formed, which can outpower the radiative output of the SMBH by a large factor. The presence of jets (and associated radio-bright lobes) makes these sources radio loud objects. Observations of X-ray binaries, the analysis of scaling relations of jets radio flux and their power, and the luminosity function of radio loud objects suggest that, in this limit, the efficiency of converting the rest mass of the accreting matter into jets mechanical energy can be at the level of 2-5\% \citep{2008MNRAS.383..277K,2008MNRAS.388.1011M}.      

It is convenient to represent the total power released by a black hole as a sum of two terms:
\begin{eqnarray}
P(\dot{m})=\left [\varepsilon_r(\dot{m}) +  \varepsilon_m(\dot{m}) \right ] c^2 \dot{M},
\end{eqnarray}
where $\varepsilon_r(\dot{m})$ and  $\varepsilon_m(\dot{m})$ are the efficiencies of the radiative and mechanical outputs. The above discussion suggests that the total efficiency, i.e., $\varepsilon_r(\dot{m})+\varepsilon_m(\dot{m})$ is of the order of a few $10^{-2}$ for a wide range of $\dot{m}$ values. However, in order to estimate the heating rate of the gas, it is necessary to specify the conversion efficiency of the released energy to heat, which turns out to be different for radiative and mechanical outputs.  As discussed in the previous section, massive black holes residing in the cores of clusters belong to the class of radio loud AGN \citep{1990AJ.....99...14B,2007MNRAS.379..894B,2007MNRAS.376.1849H,2012MNRAS.421.1569B}.  These  AGN are not very bright in the optical and X-ray bands (despite their huge mass) and, based on the weakness of lines like [OIII] relative to H$_\alpha$ in their optical spectra are classified as Low Excitation Radio Galaxies (LERG). All this suggests that SMBH in the cores of clusters  have on average strongly sub-Eddington accretion rates, placing them into a class of radiatively inefficient accretors. This means that the limit of small values of $\dot{m}$, when $\varepsilon_m(\dot{m})\gg \varepsilon_r(\dot{m})$ is more important. This however, does not exclude a possibility that they spend a small fraction of time in a regime when the accretion rate is high resulting in a radiatively efficient accretion and a powerful QSO-type AGN.  It is therefore interesting to consider both the radiatively efficient and radiatively inefficient (but mechanically efficient) limits. We write the combined heating rate $H$ as
\begin{eqnarray}
H=\left [ \alpha_r\varepsilon_r(\dot{m}) +  \alpha_m\varepsilon_m(\dot{m}) \right ] c^2 \dot{M}=\left [ \alpha_r\varepsilon_r(\dot{m}) +  \alpha_m\varepsilon_m(\dot{m}) \right ] \frac{\dot{m}}{\varepsilon_0} \times L_{\rm Edd},
\label{eq:hrate}
\end{eqnarray}
where $\alpha_r$ and $\alpha_m$ are the conversion efficiencies of radiative and mechanical energies to heat.



\begin{figure} 
\includegraphics[trim=0cm 5cm 0cm 2cm,width=0.9\columnwidth]{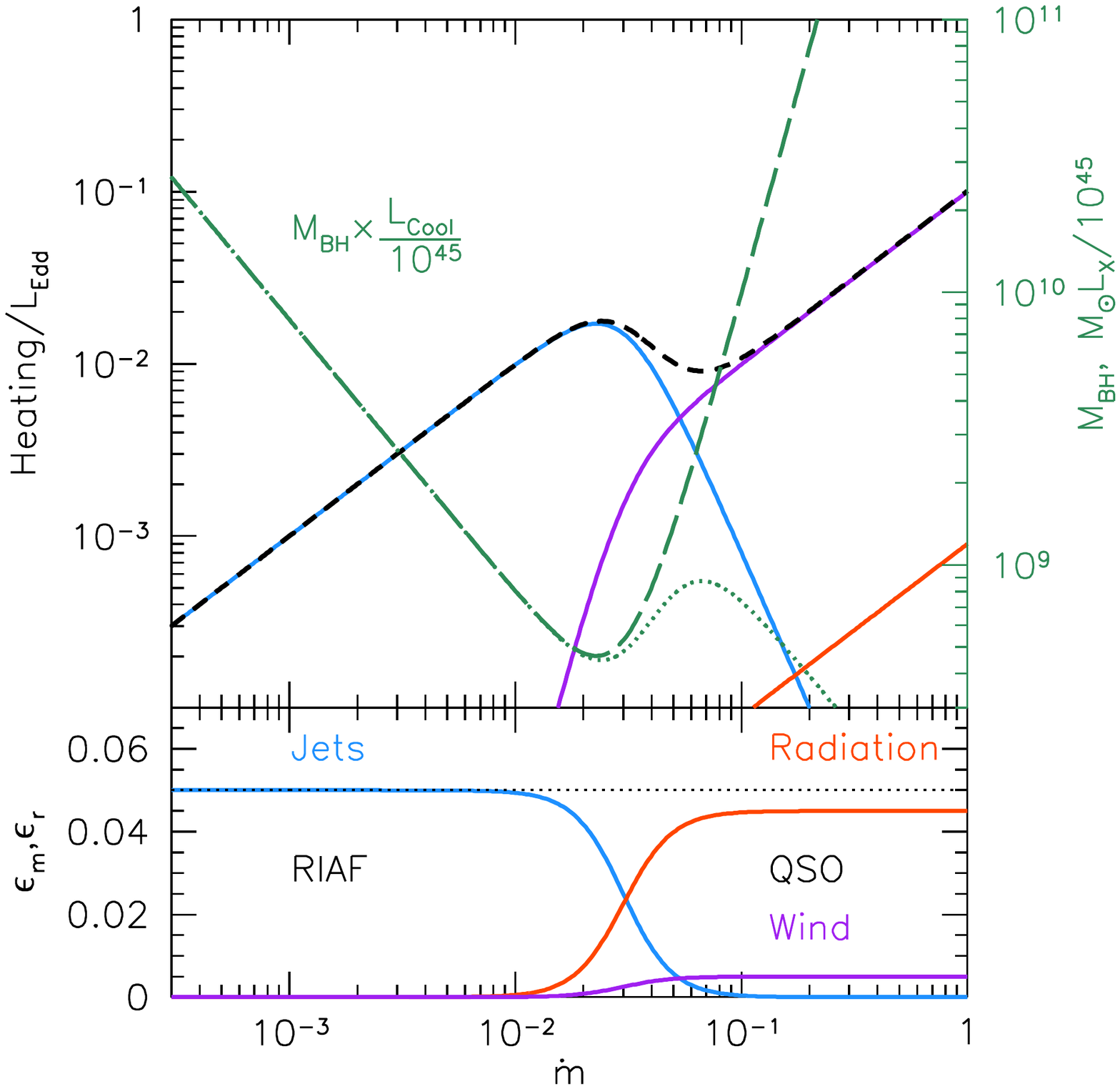}
\caption{Illustration of a heating rate in a simplified model, where mechanical energy release (jets; blue color) dominates at low accretion rates, while at high accretion rates the black hole radiative output (red) is more important \citep{2005MNRAS.363L..91C,2008MNRAS.383..277K,2008MNRAS.388.1011M}. The bottom panel shows the efficiencies $\varepsilon_r$, $\varepsilon_m$ of converting the rest mass of accreted matter into radiation  and mechanical energy (jets and winds). It is assumed that the sum of all components yields the same efficiency of $\sim 5\times 10^{-2}$ at any $\dot{m}$ (dotted line). The wind component (purple) is rather arbitrarily set to 10\% of the radiative output. 
The top panel shows the corresponding gas heating rates (see Eq.~\ref{eq:hrate}), assuming different efficiencies ($\alpha$) of converting released energy into gas heating: very high for mechanical energy $\alpha_m=1$ (jets and winds; blue and purple curves, respectively) and very low for radiation $\alpha_r=10^{-3}$ (red curve). The total heating rate is shown with the black dashed line. Finally, the dashed green line shows the mass of the black hole (see the left axis) needed to provide a heating power of $10^{45}\,{\rm erg\,s^{-1}}$ and a given Eddington ratio. This plot demonstrates that for the adopted parameters, the BH radiative output makes subdominant contribution to the direct ICM heating, while the heating by mechanical energy can be powerful even for strongly sub-Eddington objects. While on the quantitative level, the adopted coefficients can be wrong by a significant factor, this plot qualitatively illustrates that at low accretion rates it is possible to offset cooling losses without the need of an extremely massive and/or extremely fast-growing black hole. This regime is  especially relevant for clusters and groups of galaxies.}
\label{fig:feedback_mdot}
\end{figure}

\subsection{Heating efficiency by radiation} 
Consider first a hypothetical case  of a radio-quiet QSO at the center of a  rich cluster, i.e. an AGN in a radiatively efficient mode when $\dot{m}$ approaches unity. In this case, its luminosity can reach    $10^{47}-10^{48}\,{\rm erg\,s^{-1}}$, i.e. much larger than the cooling losses of the most massive clusters. What impact would this powerful radiation have on the ICM? Since the ICM is fully ionized (except for He- and H-like ions of heavy elements) and has negligible free-free opacity, the most important coupling mechanism between the radiation and gas is Compton scattering. Radiation can both cool the gas or heat the gas, depending on the gas temperature and the shape of the radiation spectrum. In other words, both Compton scattering (electrons gain energy) and inverse Compton scattering (electrons lose energy) play a role. For a typical QSO spectrum  (and in the absence of any other sources of heating and cooling), radiation will drive the temperature of the medium to the ``Compton temperature'' $T_{\rm C}\sim 2\times 10^7\, {\rm K}$ \citep{2004MNRAS.347..144S,2005MNRAS.358..168S}. 
While this value looks promising, the efficiency of coupling the radiation with matter is small, because it is set by the product of the ICM Thomson optical depth $\tau_{\rm T}\lesssim 10^{-2}-10^{-3}$ and another small factor of order $\frac{kT_{\rm C}}{m_{\rm e}c^2}\sim 4\times 10^{-3}$, which limits the amount energy that can be transferred to an electron in one scattering. This means that only a small fraction $\alpha_r\lesssim 10^{-5} - 10^{-4}$ of the black hole radiative output can go into the gas heating. While on scales of galaxies (hosting a powerful QSO) this process can be important \citep{2009ApJ...699...89C}, for rich galaxy clusters this is clearly not sufficient to prevent gas cooling. Indeed, the gas heating rate by radiation 
\begin{eqnarray}
H_r=\alpha_r\varepsilon_r(\dot{m}) c^2 \dot{M}\lesssim 10^{-4} L_{\rm Edd}.
\end{eqnarray}
For luminous clusters, this is hardly a solution. Therefore, one needs time-averaged QSO luminosity $L_{\rm R}\sim L_{\rm X}/\alpha_r\gtrsim  10^{48}\,{\rm erg\,s^{-1}}$ in every cluster with cooling losses $L_{\rm X}\gtrsim 10^{44}\,{\rm erg\,s^{-1}}$, which is clearly not observed. The low duty cycle arguments do not work in this case, since the required luminosity $L_{\rm R}$ during the active period would be even higher and exceed the Eddington limit even for the $10^{10}\,M_\odot$ BH. One can conclude that heating the gas by radiation is not a viable option for rich clusters. This does not exclude the possibility that in the vicinity of the black hole radiation can couple with the matter more strongly and drives winds/outflows that in turn interact with the gas \citep{1999MNRAS.308L..39F,2000ApJ...543..686P,2015ARA&A..53..115K}. However, AGN radiation is certainly not the ``agent'' that transports energy from the AGN to the bulk of the cooling gas. Yet another problem of keeping $\dot{m}$ close to unity is the extremely rapid growth of the black hole mass with the e-folding time $t_{\rm e}=t_{\rm s}/\dot{m}$, where $t_{\rm s}=\frac{\varepsilon_0 \sigma_{\rm T} c}{4\pi G m_{\rm p}}\sim 5\times 10^7\,{\rm yr}$ is the Salpeter time. Clearly, maintaining $\dot{m}\sim 0.1$ for, e.g., 2~Gyr would lead to a $\sim$50-fold increase of the BH mass.

In the radiatively inefficient regime, the spectrum is harder and the Compton temperature can be higher \citep{2009ApJ...691...98Y}. However, the accretion radiative efficiency and the accretion rate itself are small in this limit, which means that total heating rate will be small anyway. This is the reason, why radio mode feedback, as opposed to quasar mode feedback, is currently the favorite feedback model in application to massive groups and clusters.          

\subsection{Heating efficiency by mechanical energy}
The key feature of the radio mode AGN feedback models is the high efficiency $\alpha_m\sim 1$ of converting the energy released by the BH into gas heating (we return to this point below), i.e.
\begin{eqnarray}
H_m=\alpha_m\varepsilon_{m}(\dot{m}) c^2 \dot{M} \approx \varepsilon_{m}(\dot{m}) c^2 \dot{M}\lesssim \dot{m} L_{\rm Edd}.
\end{eqnarray}
If this supposition is correct, then the total efficiency of radio mode feedback can be much higher than for quasar mode, even in the radiatively inefficient mode, when $\dot{m}$ is small and jets dominate. Indeed, this mode prevails for $\dot{m}\lesssim \dot{m}_{\rm crit}\sim 10^{-2}$, implying that the maximum heating rate that a ``low luminosity'' AGN can provide via jets is $\sim \dot{m}_{\rm crit}L_{\rm Edd}\sim 10^{-2}\,L_{\rm Edd}$, that is about two orders of magnitude larger than the heating by radiation by the same AGN at any accretion rate. 
In other words, the total accreted mass overr the Hubble time needed to offset gas cooling, which is $\dot{M}t_{\rm H}\propto \frac{L_{\rm X}}{\alpha\varepsilon(\dot{m})}$ is smaller for the mechanical feedback than for the radiative one. However, the black hole must be sufficiently massive, so that $\dot{m}\propto \frac{L_{\rm X}}{M_{\rm BH}}$ is sufficiently small and, therefore, most of the energy is indeed released in the mechanical form.     

The assumption of $\alpha_m\sim 1$ is very natural for the majority of the mechanical feedback models. It is possible, however, to imagine cases, when this assumption is not valid. For example, when AGN jets are highly collimated, they are able to pierce through the gaseous atmosphere without much energy deposition on the way \citep{2006ApJ...645...83V}. Jets precession (known as the ``dentist's drill model'') or natural interactions of the jet with the ICM \citep[e.g.][]{2006MNRAS.373L..65H} can effectively enlarge jets' opening angle and increase the coupling efficiency. This area is a field of active research. Another example is the diffusion of relativistic protons through the ICM in a test-particle limit. In this limit, the particle can escape the halo atmosphere without much loss, although in application to real clusters it is unlikely that this regime is particularly interesting. 

\subsection{Variants of the mechanical feedback models}
Many flavors of the mechanical feedback models (forms of the energy release, transport of energy through the ICM atmosphere, and routes of energy dissipation) exist. These models are motivated by observations described in Section~\ref{sec:obs} and are trying to complement them with different sets of assumptions on the still poorly known jets and ICM microphysics, time evolution, etc. Among the most discussed scenarios are
\begin{itemize}
\item bubbles filled with cosmic rays or very hot plasma, that are thermally isolated from the ICM and interact with the ambient gas purely mechanically.  
\item outflows of hot (thermal plasma) that can mix with ICM and share their entropy with the cooling gas 
\item shock waves that are associated with sporadic outbursts of BH hole activity 
\item sound waves that are generated by expanding bubbles 
\item cosmic rays that escape from the bubbles and stream through the ICM. During this process, waves are excited in the ICM and the energy is transferred to the gas.
\end{itemize}

\subsubsection{Buoyantly rising bubbles}

\begin{figure} 
\includegraphics[trim=0cm 0cm 0cm 0cm,width=1\columnwidth]{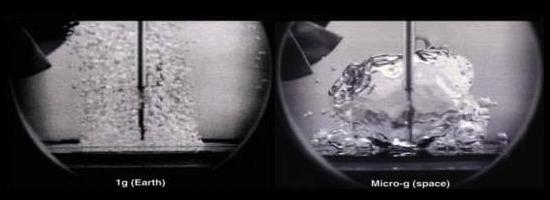}
\caption{Snapshots of NASA's boiling fluid experiment \url{https://www.nasa.gov/audience/foreducators/microgravity/multimedia/me-boiling.html}, illustrating the effects of stratification/buoyancy on the size of the bubbles. Left: boiling experiment in the presence of gravity (in the lab). Right: same boiling experiment in space (on the International Space Station). The rate of heating is the same in both cases. When there is no gravity, bubble can grow with time to a very large size. In the presence of stratification, buoyancy force pulls the bubbles up and prevents their growth leading to a sequence of bubbles with approximately the same size (for a given heating power). Larger heating power is expected to produce larger bubbles. Similar arguments are used in to infer AGN mechanical power in application to buoyant bubbles in cluster cores.}
\label{fig:boiling}
\end{figure}

The essential elements of the mechanical feedback model can be most easily illustrated using the scenario involving bubbles. As stated earlier, an AGN releases mechanical energy (in a form of a jet or outflow) at a steady rate and inflates an X-ray cavity in the  ICM, filled by cosmic rays or very hot plasma. Observationally (see \S~\ref{obscase}), they appear as depressions in X-ray surface brightness and as bright bubbles in radio band, clearly powered by synchrotron emission of relativistic electrons (and, possibly, positrons). The energy associated with the expanding bubble can be calculated as the change of the total gas energy $\displaystyle E_{\rm tot}=\int \left [P/(\gamma-1) + \rho v^2/2+ \rho\phi \right ] dV$, i.e. the sum of thermal, kinetic and potential energies integrated over the volume. Here $\phi$ is the gravitational potential. 
The initial (relatively short) expansion phase, when the bubble is small, is supersonic and the boundary of the expanding bubble drives a strong shock into the ICM \citep{1998ApJ...501..126H,2001ApJ...549L.179R}. For the strong shock, the gas is compressed by a factor of 4 (for the ICM adiabatic index $\gamma=5/3$) and the shock stays close to the bubble boundary. Most of the energy during this phase goes into kinetic energy and shock heating of the gas. As the size of the bubble increases, the expansion rate slows down and eventually becomes subsonic. Accordingly, the shock weakens, moves away from the bubble boundary and becomes a sound wave. During this stage, the pressure inside the bubble is approximately the same as the ICM pressure and, therefore, the kinetic energy and the change of its thermal energy of the ICM can be neglected.  In this case, the total energy, deposited by the AGN into the inflation of the bubble can be straightforwardly estimated as the enthalpy of the bubble $H=\frac{\gamma_{\rm b}}{\gamma_{\rm b}-1}PV_{\rm b}$, i.e. the sum of the internal energy of the bubble and the work done on the ICM (at constant pressure $P=P_{\rm ICM}$),
where, $V_{\rm b}$ is the volume of the bubble, $P=P_{\rm ICM}$ is the pressure inside the bubble. The $PV_{\rm b}$ can be viewed as potential energy of the bubble, created by displacing the gas to larger radii. This approximate expression is used when estimating the energy of the observed bubbles, see Section~\ref{obsenergetics} and Eq.~\ref{eq:ecav}.   The subsonic phase continues with ever decreasing expansion velocity, until buoyancy effects start to play a role \citep{2000A&A...356..788C}. Indeed, the cluster atmosphere is stratified and pressure gradients acting on the bubble  filled with very hot plasma (and, therefore, underdense compared to ICM) will drive the bubbles away from the center. The importance of buoyancy for bubbles of relativistic plasma has been already recognized by \citep{1973Natur.244...80G}. Effectively, buoyancy introduces an additional timescale $\displaystyle t_{\rm b}\sim \sqrt{l_{\rm b}/g}$ to the problem and, once the time scale associated with the bubble expansion becomes longer, the bubble deforms/breaks and starts rising in the potential well of the cluster \citep{2001ApJ...554..261C,2002MNRAS.332..271R}. Here  $l_{\rm b}$ and $g$ are the size of the bubble and gravitational acceleration respectively. The whole process of bubbles' formation and evolution bears a strong similarity to a familiar example of boiling water in a pot, as illustrated in Fig.~\ref{fig:boiling}. Together with the expression \ref{eq:ecav} for the total energy associated with the bubble this gives an estimate of the AGN power $\displaystyle L_{\rm AGN} \sim H/t_{\rm b}$ \citep{2000A&A...356..788C,2000ApJ...534L.135M}. Importantly, the buoyancy becomes important when the expansion velocity of the bubble is already subsonic (or at most transsonic). 
This validates the use of Eq.~\ref{eq:ecav}. Several variations of the above estimate have been considered (see Section~\ref{obsenergetics}), but most of them obey the same scaling and differ by a numerical factor of a few. 

Once the bubble is detached from the jet, its evolution is controlled by the buoyancy that pulls the bubble up and by the drag force from the ambient gas that brakes the bubble motion. For very light bubble (almost massless), it quickly attains a terminal velocity when the buoyancy force is balanced by the drag. Namely,
\begin{eqnarray}
F_{\rm drag}\approx  -F_{\rm buoyancy}=g\rho_{\rm ICM}V_{\rm b},
\label{eq:fdrag}
\end{eqnarray}
where $\rho_{\rm ICM}$ is the mass density of the ICM. Accordingly, $-F_{\rm drag}$ is the force acting on the ICM. When the bubble with fixed volume raises along the radius by  $\Delta r$, the work done by the drag force  $F_{\rm drag} dr\approx  g\rho_{\rm gas}V_{\rm b}\Delta r$ is equal to the change of the bubble potential energy. However, since the bubble stays in approximate pressure equilibrium with the ambient gas, its volume gradually increases, effectively converting bubble's thermal energy into potential energy of the displaced gas. This process guarantees that all the energy associated with the bubble  will be released to the ambient thermal gas \citep{2001ApJ...554..261C,2002MNRAS.332..729C,2001ASPC..240..363B}. For $\gamma_{\rm b} = 4/3$ (bubble filled with relativistic particles), the bubble transfers half of the available energy to the ICM after crossing $\approx 2.8$ pressure scale heights of the atmosphere \citep{2017ApJ...844..122F}. For $\gamma_{\rm b} = 5/3$ (bubble filled with thermal plasma), this will happen after  less than 2 scale heights.

\begin{figure}[!ht]
\includegraphics[trim=0cm 0cm 0cm 0cm,width=1\columnwidth]{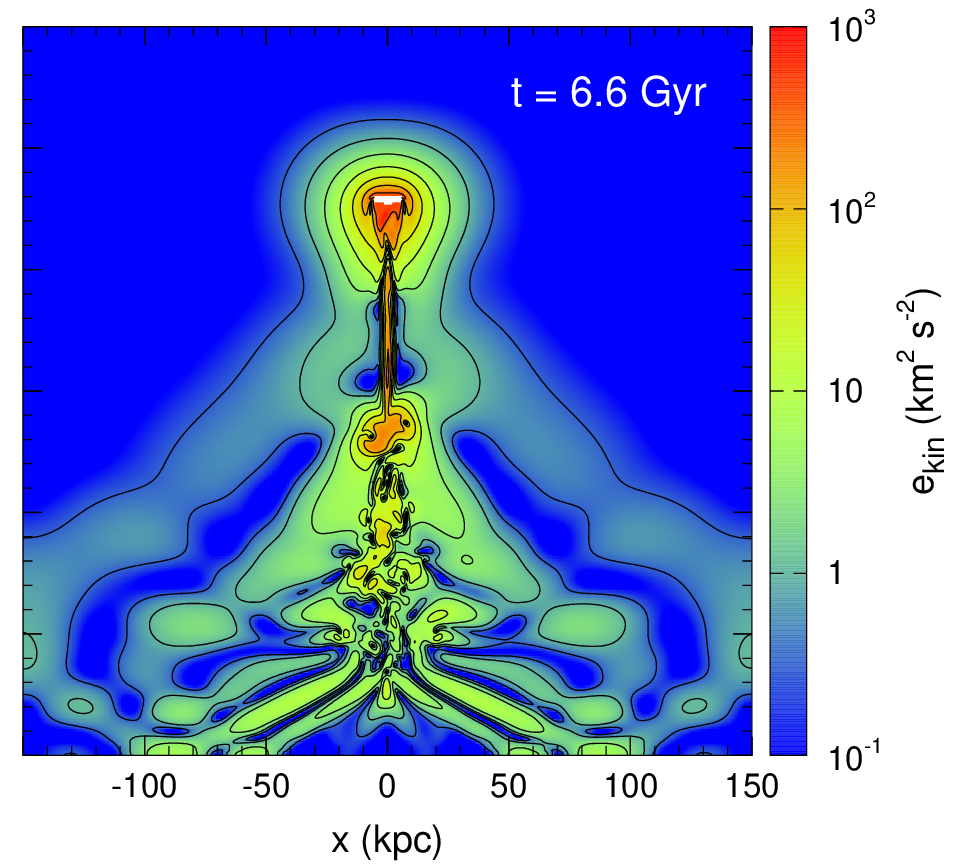}
\caption{Excitation of internal waves by a bubble rising in a stratified atmosphere \citep{2018MNRAS.478.4785Z}, illustrating one of the (many) channels of extracting energy from the bubble. Essentially independently on the physical mechanism of the ``friction'' between the bubble and the ICM, most of the energy, initially stored as the enthalpy of the bubble will be transferred to the ambient gas when the bubble crosses a few pressure scale heights of the atmosphere. However, the rise velocity and the type of perturbations induced in the ICM (e.g., properties of the gas velocity field) do depend on the physical processes dominating the gas/bubble interaction. In this particular model, bubbles transport energy in vertical direction, while internal waves spread it in lateral directions.}
\label{fig:iw}
\end{figure}

This conclusion is essentially independent of the nature of the drag force, which could be due to viscosity, driving turbulence, excitation of waves in the ICM (see Fig.~\ref{fig:iw}), entrainment of low entropy gas by the rising bubble or something else. In this model, the bubbles themselves are responsible for the transportation of energy in the radial direction, while internal waves transport energy in the azimuthal direction. In any case, the basic conclusion that the coupling efficiency of the buoyantly rising bubbles to the gaseous atmosphere is close to $100\%$ holds, since it is governed by energy conservation. 

The energy extracted from the bubbles can be first stored in different forms (and even undergo a number of transformations) before it is converted into heat. What are these intermediate repositories of energy? 
The answer to this question sensitively depends on the properties of the ICM, such as transport coefficients, strength, and topology of the magnetic field, etc., as well as on the properties of the bubbles. As a result, no firm consensus has been reached on the relative contributions of these effects to the energy extraction process. For instance, one of the possible routes is the excitation of internal waves by rising bubbles that spread the energy through the cluster core, producing anisotropic gas motions (turbulence), which then dissipate. The turbulence dissipation is invoked in several scenarios, \citep[e.g.][]{2005ApJ...622..205D,2005MNRAS.359.1041R,2006MNRAS.372.1840R,2014Natur.515...85Z} and can be tested with observations, once accurate gas velocity measurements become available. In the model of \cite{2018MNRAS.478.4785Z}, the bubbles were treated as rigid bodies rising in unmagnetized fluid. If instead, they are able to mix with the ISM, the excitation of turbulence can become less efficient \citep[e.g.][]{2015ApJ...815...41R,2016ApJ...829...90Y,2018ApJ...857...84B}. Magnetic fields could partly stabilize bubbles and, also, induce a drag on the rising bubble \citep{2006MNRAS.373...73L}, which can be larger than the pure hydrodynamic drag \citep{2008ApJ...677..993D}.

\subsubsection{Winds, outflows of thermal plasma, and mixing}
Even a simpler solution is possible if the AGN is releasing energy as thermal winds/outflows, especially, if they are not strongly collimated \citep[e.g.][]{2002Natur.418..301B,2004MNRAS.348.1105O,2005ApJ...632..821P,2010MNRAS.408..961P,2006ApJ...643..120B,2006MNRAS.366..397S,2016MNRAS.455.2139H}. Unlike relativistic protons, which exchange energy with thermal protons via collisions very slowly, the thermal plasma of the wind can easily share its energy with the cooling gas by mixing and subsequent fast energy exchange between hotter and colder particles. As with other flavors of the mechanical feedback, the coupling efficiency of  these outflows with the ICM is expected to be very high. Besides, from the point of view of numerical simulations, this case is especially transparent, and can be modeled in the frame of conventional hydrodynamics. The main issue is whether the majority of SMBHs in cores of clusters and groups produce such winds, which are more typical for high accretion rates.

\subsubsection{Strong shocks}

\begin{figure} 
\includegraphics[trim=0cm 0cm 0cm 0cm,width=1\columnwidth]{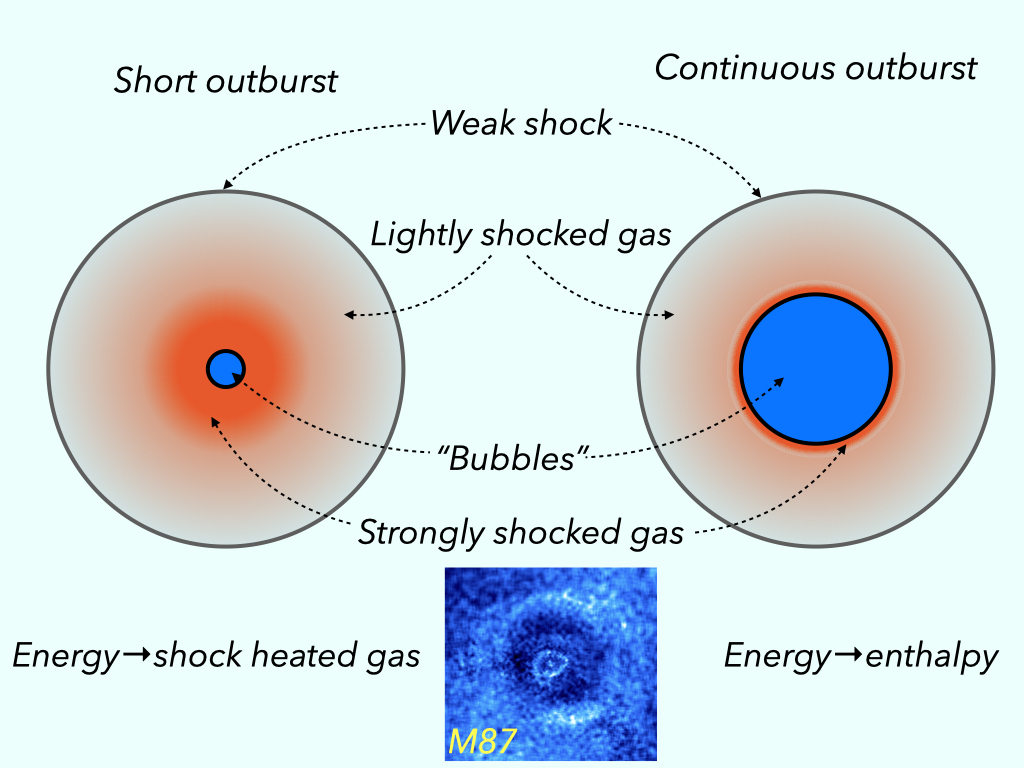}
\caption{Qualitative comparison of a very short (left) and continuous (right) spherically symmetric outbursts with the same total energy in a homogeneous gas \citep[following][]{2017ApJ...844..122F,2017MNRAS.468.3516T}. For the short outburst, the AGN instantaneously releases all energy in a form of highly-overpressurised plasma (marked in blue).  This gas drives strong shock in the ICM and most of the energy goes into acceleration and shock heating of the ICM. At later stages, the shock becomes weak (this stage is shown in the figure), essentially a sound wave, carrying 10-15\% of the total energy, leaving behind a small region of filled with the AGN-injected plasma and a large region of the shock-heated gas. For the continuous outburst (right), the AGN gradually releases the same amount of energy. In this case, strong shock is present only during the very early stage of the energy release. As a result, not much shock heated gas is present in the system and most of the energy goes into bubble enthalpy (big blue circle). In fact, the bubbles might appear surrounded by cool rims, arising due to adiabatic expansion of the gas. Real objects, of course, are much more complicated than depicted in the figure.
The \textit{Chandra} X-ray image of M87 (3.5-7.5 keV; flat-fielded) is shown as an example of an object where we do see clearly a weak shock (outer circle) and AGN-injected gas (inner oval-shaped structure). The analysis of the M87 case suggests that the short outburst scenario does not fit the data well \citep{2017ApJ...844..122F}. }
\label{fig:shock_sketch}
\end{figure}

Strong shocks provide a straightforward way of heating the ICM \citep[e.g.][]{1998ApJ...501..126H,2017ApJ...847..106L}. Under certain conditions, this might be the dominant heating mechanism. Strong shocks in the ICM could be created by collimated outflows or by time variable energy release. For instance, if the AGN spends most of the time in a quiescent state and releases energy in a rare, short, and very powerful outburst, most of the energy will go into the shock heating of the ambient gas.  This is in contrast to the continuous/slow energy injection discussed in the rising bubbles scenario, when most of the energy goes into the enthalpy of the inflated bubble\footnote{A strong shock might be present during the initial phase of the bubble inflation, but this accounts for a small fraction of the released energy} and heating takes place during the rise of the bubble. 
The outcome of an instantaneous outburst (see Fig.~\ref{fig:shock_sketch}) is very different: a tiny bubble filled with AGN ejecta surrounded by a large envelope of the shock-heated ICM\footnote{Note that the time interval between successive outbursts should be long enough, so that the shock-heated gas can leave the cluster core. Otherwise, the continuous injection scenario is more appropriate.}. This envelope can further mix with a larger mass of gas, providing efficient distribution of energy over the cooling gas, or part of it could rise buoyantly, repeating the rising bubble story. Either way, the efficiency of energy thermalization is expected to be high. 

\subsubsection{Sound waves}
Another model, motivated in particular by a sequence of ripples seen in the X-ray images of the Perseus cluster,  associates ICM heating process with weak shocks and sound waves \citep{2003MNRAS.344L..43F,2006MNRAS.366..417F,2004ApJ...611..158R,2009ApJ...699..525S,2009MNRAS.395..228S}. Similar ripples were also found in Abell~2052 \citep{2011ApJ...737...99B} and the Centaurus cluster \citep{2016MNRAS.457...82S}.
In this model, the activity of the central black hole inflates bubbles and generates weak shocks and sound waves, which propagate radially and act as carriers of energy through the cooling gas. An attractive feature of this model is the isotropic transport of energy in all radial directions. The rate of waves dissipation (= heating of the gas) depends on the not yet fully understood effective transport coefficients and on the waves amplitude in a weakly collisional cluster plasma \citep{2018ApJ...858....5Z,2020JPlPh..86f9003K}. Yet another channel of tapping energy from sound waves and shocks was discussed in \cite{2005ApJ...634L.141H,2012ApJ...746..112F}. Namely, when the waves propagate through the ICM filled with bubbles of relativistic plasma, a vortex velocity field is formed around the bubbles
which can subsequently be dissipated. 

Another open question is the efficiency of waves generation by the AGN-inflated bubbles. For a spherically symmetric expanding bubble in a homogeneous medium, less than 13\% of the total energy can go into sound waves \citep{2017MNRAS.468.3516T}, although for collimated flows this fraction can be larger \citep{2019ApJ...886...78B}. 

\subsubsection{Heating by Cosmic rays streaming}
The ``rising-bubbles'' model does not depend explicitly on the content of the bubbles. It only assumes that bubbles are buoyant with respect to the ambient ICM, i.e. they are much lighter than the ICM at the same pressure.  Observed radio emission from bubbles shows that relativistic electrons are present inside the bubbles. It is very common in astrophysical conditions that the energy density of relativistic protons dominates that of the electrons by a large factor. It is, therefore, reasonable to assume that relativistic protons make most of bubbles' pressure\footnote{The lack of SZ signal from large bubbles in the  MS~0735.6+7421 cluster \citep{2019ApJ...871..195A} is consistent with this assumption, but does not yet exclude the possibility that bubbles are filled with thermal but very hot plasma. In the latter case, the temperature of the plasma must be higher than hundreds of keV.}. If cosmic ray protons (CR) can escape from bubbles and mix with the ICM, then yet another channel for ICM heating appears \citep[e.g.][]{1991ApJ...377..392L,2008MNRAS.384..251G,2013MNRAS.434.2209W,2013ApJ...779...10P,2017ApJ...844...13R,2018MNRAS.481.2878E}. Direct interactions of CRs with thermal protons via hadronic collisions are not very effective \cite[e.g.][]{1996SSRv...75..279V}, however, CRs might stream through the ICM (move with respect to the gas frame) driven by the gradient CRs pressure. In this case, CRs streaming with super-alfvenic speed can collectively excite waves in plasma, which scatter cosmic rays limiting their streaming velocity. Dumping of these plasma waves provides the heating channel for the gas. The model predicts the volumetric heating rate $h_{\rm CR}\sim |v_A\nabla P_c|$, where $v_A$ and $\nabla P_c$ are the Alfven velocity and gradient of the CR pressure, respectively. Note that for the buoyant bubbles scenario one can write the heating rate (that follows from eq.~\ref{eq:fdrag}) in a similar form, namely $h_{\rm bub}\sim |f_b v_{\rm rise}\nabla P_t|$, where $f_b$, $v_{\rm rise}$ and $\nabla P_t$ are the fraction of volume occupied by bubbles, their rise velocity and the gradient of the total pressure in the atmosphere, respectively. It is  interesting to compare the heating rates in the cluster core:
\begin{eqnarray}
\frac{h_{\rm bub}}{h_{\rm CR}}\sim \frac{v_{\rm rise} }{v_A}\frac{f_b\nabla P_t}{\nabla P_c}\sim \frac{v_{\rm rise} }{v_A}.
\end{eqnarray}
In realistic conditions this ratio can be larger than unity, implying that when cosmic rays are confined inside the bubble, smaller amount of CRs is needed at any given moment to provide the same level of heating. Nevertheless, the final efficiency of extracting energy from CRs is expected to be high for both processes.

\subsubsection{Broader outlook} 
Despite the diversity of suggested mechanical feedback models, they all share the common property - high coupling efficiency of the generated energy with the gas. This ensures that all released mechanical energy is used for gas heating. This is the fundamental reason why all these models appear to be successful in solving the main problem of compensating for the gas radiative losses and preventing massive star formation in galaxy clusters. There are three implications: (a) it is likely that the mechanical feedback can (and does) offset cooling losses in clusters and groups, (b) it is possible that several processes are contributing to heating\footnote{Once again, we can use the analogy with dissipation of fluid motions, when in the steady-state the heating rate is equal to the driving power and does not depend on viscosity or other characteristics of the fluid},  and (c) more careful examination of the data is needed to understand, which processes dominate (not necessarily the same in all clusters and groups). The hope is that gas velocity measurements (amplitudes, spatial correlation length, anisotropy) by forthcoming generation of cryogenic X-ray bolometers like \textit{XRISM} \citep{2022arXiv220205399X}, \textit{Athena} \citep{2013arXiv1306.2307N}, \textit{LYNX} \citep{2018arXiv180909642T} will become instrumental in clarifying this question. Currently, only indirect probes of the velocity field are possible, except for a short observation of the Perseus cluster with \textit{Hitomi} \citep{2016Natur.535..117H}. Another important piece of information could come from sensitive gamma-ray data that would provide constraints on relativistic protons mixed with the ICM, accompanied by sensitive low-frequency radio surveys.

\subsubsection{Cooling of the gas}
One of the actively debated questions is the origin of the dense and cool gas, which is often found in the cores of clusters and groups almost every time when the hot gas cooling time is shorter than $\sim 10^9\,{\rm yr}$. This cool ($\sim 10^4\,{\rm K}$) gas, most easily seen in ${\rm H}_\alpha+{\rm NII}$ lines\footnote{There is molecular gas too, which is bright in, e.g., CO lines.}  often forms filaments extending over tens of kpc. It has been long recognized that this gas cannot be simply the cooling of the hot X-ray-emitting gas through the temperature $T_{\rm cool}\sim 10^4\,{\rm K}$. In this case, the ratio of the cool gas luminosity to the X-ray gas cooling losses would be $\sim T_{\rm cool}/T_{\rm hot}\sim 10^{-3}$. The estimated luminosities of the cool gas are much higher, implying that an extra source of energy is needed to power the cool gas. Once the presence of the energy source is assumed, the ``age'' of the cool gas becomes uncertain. It could be formed relatively recently or could be very old. It all depends on the availability of the energy source. To some extent, this problem is similar to the cooling of the hot gas, although the formally calculated cooling time of the $\sim 10^4\,{\rm K}$ gas is orders of magnitude shorter. It is plausible that AGN provides the power for the filaments too, although there is no consensus on the exact physical mechanism that supplies the energy to the cool gas. The options discussed include thermal conduction, shocks, turbulence, hot particles penetrating the filaments, magnetic reconnection, and photoionization.  Numerical simulations are challenging too because of the large scale and time separations between the hot and cold phases as well as uncertainties in a proper description of the stratified hot gas affected by the AGN feedback. We discuss this problem further in Section~\ref{sec:sim1} in particular in relation with the non-trivial question of the stability of stratified cooling atmospheres.








\section{Simulating AGN feedback in general: basic models and important parameters}\label{sec:sim1}

AGN feedback is a crucial ingredient in large-scale cosmological simulations to reproduce the quenched population at $z=0$. In galaxy-scale simulations, AGN feedback is the key to keeping massive galaxies and galaxy clusters quiescent, preventing the development of a cooling catastrophe. 

Modeling AGN feedback is rather complicated and cannot be done from first principle. For one thing, many aspects of the accretion physics and feedback processes are still poorly understood. For another, SMBH accretion and jet/radiation production happen near the event horizon, typically at $\sim$ AU scale, many orders of magnitude smaller than the scale of a galaxy. Due to the resolution limit, cosmological simulations and galaxy-scale simulations cannot resolve even close to the event horizon. Therefore, all numerical simulations rely on ``sub-grid'' models to mimic the real accretion and feedback processes.

The amount of energy released from SMBH accretion is proportional to the accretion rate: $\dot{E}_{\rm BH}=\epsilon \dot{M}_{\rm BH}c^2$, where $\epsilon$ is the feedback efficiency, which is different from the accretion efficiency $\eta$. $\epsilon$ can be smaller than $\eta$ because some of the energy extracted from accretion can be ``wasted'' (e.g., in the form of radiation in quasars) without causing damage to the host galaxy. Numerical models are only concerned with the $\dot{E}_{\rm BH}$ that is participating in active feedback. 

Different sub-grid models of AGN feedback can differ both in the amount of energy injected ($\epsilon$) and in the way energy is injected. Many models also make some distinction between the low accretion rate mode (hereafter low-state mode) and the high accretion rate mode (hereafter high-state mode), but exactly how it is done can vary from model to model. Most simulations also impose an Eddington limit at the high rate end. Different simulations or different authors describing the same simulations may also refer to the feedback modes differently. Indeed, the low-state mode is sometimes called the ``radio mode'' or ``jet mode'' based on the observed phenomena associated with this mode of accretion, but can also be called ``kinetic mode'' which is how it is often modeled in simulations (see also Figure~\ref{fig:feedback_modes} for a summary of accretion and feedback modes discussed in this Chapter).

\subsection{Modeling AGN feedback in cosmological simulations}

Although AGN feedback in the real universe involves many complicated physical processes, in existing cosmological simulations, it is typically modeled with only simple hydrodynamic ingredients. That is, feedback energy is injected as kinetic or thermal energy. Relativistic jets, cosmic rays, and magnetic fields are usually ignored. When radiation is included, it is usually modeled as a simple heating term. Still, even within the pure hydro regime, there are many bells and whistles in modeling AGN feedback. Below we list some of the models that have been used in cosmological simulations over the past decade.

AGN feedback in Illustris \citep{2015MNRAS.452..575S} is a modified version of \citet{2005MNRAS.361..776S}, and is mainly modeled with two modes separated at an Eddington ratio (accretion rate) of $f_{\rm Edd} = 0.05$. In both the high-state mode (referred to as the ``quasar'' mode in Illustris) and the low-state mode (referred to as the ``radio'' mode), feedback energy is injected as thermal energy. The main difference is that in the low-state mode, thermal energy is injected into large bubbles with a radius of 100 kpc randomly placed around each black hole. Feedback efficiency is set to be $\epsilon = 0.05\eta$ and $\epsilon = 0.35\eta$ for quasar and radio mode, respectively, where $\eta=0.1$. A similar feedback model is also used in the Magneticum simulations with slightly different parameters \citep{2014MNRAS.442.2304H}. Close to the Eddington limit, an additional ``radiative'' feedback is added by modifying the net cooling rate of the gas surrounding the black holes.

In IllustrisTNG (TNG), the high-state mode is kept the same as Illustris with thermal energy injection, but both the low-state mode and the transition are modified \citep{2018MNRAS.479.4056W}. In TNG, low-state mode feedback is modeled with kinetic energy injection in random directions from the black hole. TNG models the transition between the two modes in a rather unique way. Instead of using a fixed $f_{\rm Edd}$ like most other simulations, $f_{\rm Edd}$ in TNG is also a function of $M_{\rm BH}$, effectively making more massive SMBHs enter the low-state mode more easily. 

Horizon-AGN separates the two feedback modes at $f_{\rm Edd} = 0.01$ \citep{2012MNRAS.420.2662D, 2016MNRAS.463.3948D}. In the high state, thermal energy is injected isotropically into a small sphere around the SMBH with $\epsilon = 0.015$. In the low state, feedback is modeled as a pair of jets (high-velocity bipolar outflows) with $\epsilon = 0.1$. 

For the SIMBA simulation \citep{2019MNRAS.486.2827D}, the two feedback modes are separated at $f_{\rm Edd} = 0.2$. Both modes inject kinetic energy. High-state black holes produce winds with a velocity that increases with $M_{\rm BH}$. Low-state black holes produce jets whose velocity increases with decreasing $f_{\rm Edd}$, and is capped at 7000 $\rm km\, s^{-1}$ at $f_{\rm Edd}\lesssim 0.02$. They also include a $M_{\rm BH}$ limit for the jet mode and model X-ray heating broadly following \citet{Choi2012}. 

The EAGLE simulation \citep{2015MNRAS.446..521S}, unlike many of the other cosmological simulations, does not distinguish between the two modes of feedback. They model only one mode with stochastic thermal energy injection with $\epsilon = 0.015$. To prevent cooling and achieve efficient feedback, feedback only occurs when enough energy has been accumulated to raise the gas temperature above $10^{8.5}$ K. A similar scheme is also adopted in the BAHAMAS project, a suite of large-volume cosmological hydrodynamical simulations \citep{2017MNRAS.465.2936M}.

All the examples listed above are large-scale cosmological simulations with box sizes of $\sim 100\,{\rm Mpc}$ or larger. They all employ different AGN feedback models. The model parameters are usually tuned using smaller-scale simulations before the production run, with a goal to match certain observed galaxy scaling relations (e.g. the $M_{\rm BH} - M_*$ relation) at $z=0$. They are also dictated by the simulation resolution. The same model run with the same code may require different parameters when run at a different resolution. The cost of these large-scale simulations makes it difficult to conduct a thorough model variation study, but some efforts have been made within single frameworks, such as the TNG model variation study. There are also smaller cosmological simulations which focus on how the feedback models affect the simulation results \citep[e.g.,][]{2017MNRAS.464.2840A, 2017MNRAS.470.1121T, 2018ApJ...860...14B}. Comparing galaxy scaling relations across different cosmological simulations can also reveal how feedback models affect galaxy evolution \citep[e.g.,][]{2021MNRAS.503.1940H}. 

Additionally, it is worth noting that there are also zoom-in simulations focused on galaxy clusters and calibrated with cluster properties. Examples include the Dianoga simulations \citep{2020A&A...642A..37B}, the FABLES simulations \citep{2018MNRAS.479.5385H}, the Cluster-EAGLE (c-eagle) simulation project \citep{2017MNRAS.471.1088B}, and the Hydrangea simulations \citep{2017MNRAS.470.4186B}. 

What we have learned from these models is that in general, kinetic feedback (in the form of jets or winds) is more effective than thermal feedback. For example, the kinetic feedback in the low-state of TNG is more effective than the pure thermal feedback in Illustris, resulting in a more realistic quenched population at low redshift \citep{2018MNRAS.479.4056W}. Kinetic feedback also tends to produce CGM properties in better agreement with the observations while thermal feedback often leads to overly dense halos which are too bright in the X-ray \citep{2015MNRAS.449.4105C}.

Thermal feedback is widely used in cosmological simulations mainly because it is easy to implement and is relatively stable numerically compared with fast outflows. However, if thermal energy is injected into cold dense regions, it gets quickly radiated away. If thermal energy is injected into the hot gas in massive systems, it may even create an inverted entropy gradient which leads to runaway cooling \citep{2017ApJ...841..133M,2017ApJ...845...80V}. Thermal feedback can still be effective if cooling is artificially suppressed (e.g., in EAGLE), which preserves the energy and allows the development of thermally-driven outflows \citep[e.g.,][]{2017MNRAS.470.1121T}. 

Kinetic feedback does not suffer from radiative energy loss as thermal feedback. In addition, kinetic feedback can push gas to larger distances from the centers of galaxies, either directly removing cold star-forming gas, or reducing the hot gas density in the central region, which suppresses further cooling. These two effects are sometimes referred to as the  ``ejective'' and ``preventative'' feedback. It has been recently argued that sweeping gas out of the galaxy is the key to quenching, which results in a tight correlation between $M_{\rm BH}$ and the central potential well of the host galaxy \citep{2020MNRAS.493.1888T, 2020MNRAS.491.2939O, 2020ApJ...899...70V}. We note in passing that many of these simulations are aimed to explain scaling relations for galaxies rather than clusters. Therefore, not all the failures or successes directly apply to the more massive halos.

\subsection{Modeling AGN feedback in idealized simulations}

AGN feedback has also been studied extensively in idealized simulations. Typically, the setup includes a static potential based on the observations of real systems and a static black hole. Because idealized simulations focus on only one galaxy or galaxy cluster, they are able to achieve higher resolution than cosmological simulations. Their relatively low computational cost also allows the inclusion of more complicated physical processes and a more thorough model/parameter variation study. The idealized setup, though limiting in certain aspects, makes it relatively easy to analyze and understand the simulation results.

In this section, we limit our discussions to idealized simulations of massive systems, i.e., group centrals and cluster centrals. We mainly focus on two aspects of AGN feedback modeling: kinetic (jet) feedback and non-hydro processes. 

The basic equation for kinetic AGN feedback is 
\begin{equation}
\dot{E}_{\rm BH}=\epsilon \dot{M}_{\rm BH}c^2 = \frac{1}{2}\dot{M}_{\rm out}v_{\rm out}^2 \,,
\end{equation}
where $\dot{M}_{\rm out}$ and $v_{\rm out}$ are the outflow rate and velocity, respectively. In many numerical simulations, the accretion rate $\dot{M}_{\rm BH}$ is set to be a fixed fraction of the computed accretion rate which is sometimes taken as the inflow rate through the inner boundary $\dot{M}_{in}$ (see more discussions in Section~\ref{accretionYuan} on accretion modeling), so we have $\dot{M}_{\rm BH}=f\dot{M}_{in}$ and $\dot{M}_{\rm out}=(1-f)\dot{M}_{in}$ from mass conservation. $\epsilon$ and $f$ together determine the outflow velocity $v_{\rm out}$:
\begin{equation}
v_{\rm out}=\sqrt{\frac{2\epsilon f}{1-f}}c \,.
\end{equation}
Typical $v_{\rm out}$ used in idealized simulations is on the order of thousands up to a few tens of thousand km s$^{-1}$. A larger $f$ creates a faster, lighter outflow with a larger $v_{\rm out}$ and a smaller $\dot{M}_{\rm out}$, while a smaller $f$ corresponds to a slower, heavier (more mass-loaded) outflow. 

On top of the basic setup, there can be many features added within the kinetic feedback framework. Both $\dot{M}_{\rm out}$ and $v_{\rm out}$ can follow a distribution function at the jet base such that each cell gets a different amount of $\dot{M}_{\rm out}$ or is assigned a different $v_{\rm out}$ based on its exact location. The ``jets'' can precess at a certain angle with a certain period, both of which are additional free parameters of the simulation. The outflow can also have an opening angle, making it a ``wind'' rather than bipolar ``jets'' with no opening angle. A fraction of the feedback energy $\dot{E}_{\rm BH}$ can also be injected as thermal energy, making the outflows hotter and higher pressured. 

It has been shown very early on that simple light hydro jets cannot solve the cooling flow problem in simulated galaxy clusters \citep{2006ApJ...645...83V}. This is because light hydro jets tend to carve a channel and deliver the energy outside of the cool core where heating is needed. Later numerical models that successfully suppress cooling flows typically use at least one of the following methods to increase the impact area of the jets: heavier (more mass-loaded) jets, precessing jets, larger opening angles, and hotter jets with both kinetic and thermal energy injection at the jet base \citep[e.g.,][]{2013MNRAS.432.3401G, Li2014b, Prasad2015, 2016ApJ...829...90Y, 2017ApJ...841..133M, 2021MNRAS.507..175S}. All these methods can be successful in balancing radiative cooling, but the gas may have different phase-space distribution, and the exact heating mechanism can also depend on the model, which we discuss in more detail in Section~\ref{sec_sim_heating}. 

In addition to pure hydrodynamic processes, many idealized simulations have also explored the effects of other physical processes on AGN feedback, such as thermal conduction, viscosity, radiation, magnetic fields, and cosmic rays. 

Thermal conduction can be an important heating mechanism in cool-core clusters if it is not strongly suppressed \citep[e.g.,][]{2002MNRAS.335L...7V, 2011ApJ...740...28V}. However, it has been theorized that anisotropic thermal conduction in cluster cool cores can cause heat-flux-driven buoyancy instabilities, re-orienting the magnetic fields azimuthally and thus preventing conductive heating from hotter ICM outside the cool core \citep{2008ApJ...673..758Q, 2009ApJ...703...96P}. \citet{2016ApJ...818..181Y} show that SMBH jets can randomize magnetic field lines to allow some radial conductive heat flux, which in turn reduces the burden on the SMBH itself.

Viscosity can be important in shaping the jet-driven bubbles observed in the X-ray. In pure hydrodynamic simulations, fluid instabilities can disrupt hot bubbles easily. It has been shown that both isotropic viscosity and anisotropic viscosity (with the right magnetic field configuration) can help maintain the shape of these bubbles \citep{2009ApJ...704.1309D, 2015ApJ...803...48G}. 

In simulations of massive systems at low redshifts, radiation feedback from the SMBH is often ignored because the SMBH is typically in the radio mode with a radiatively inefficient accretion flow. Some simulations do include the X-ray feedback as an additional heating term, similar to some cosmological simulations \citep{2014ApJ...789..150G}. \citet{2019ApJ...877...47Q, 2020NatAs...4..900Q} have modeled radiative feedback using ray-tracing radiative transfer, and although radiation does not significantly affect the global thermal balance, it does have an impact on the phases of the outflows and how the dusty cold filaments develop.

The hot gas in massive systems is only weakly collisional and magnetized with a plasma beta of $\sim 100$. However, magnetic fields are crucial in jet launching, and the jets themselves can also be highly magnetized. It is shown that magnetic jets can create X-ray cavities and the injected magnetic fields can be transported throughout the cluster and amplified by turbulence \citep{2008ApJ...681L..61X, 2011ApJ...739...77X}. Relativistic magnetic jets can also inflate cavities via kink instability \citep{2016MNRAS.461L..46T}. There are also numerical studies that do not launch magnetized jets, but the surrounding gas is magnetized \citep[e.g.,][]{2007MNRAS.378..662R, 2016ApJ...818..181Y, 2021MNRAS.504..898W}.

In terms of achieving global thermal balance, magnetized jets are not much different from pure hydro jets \citep{2021MNRAS.507..175S}. This is not surprising as magnetic fields are dynamically unimportant in weakly magnetized plasma. However, magnetic fields can impact how viscosity and conduction operate\citep{2010ApJ...720..652S}. In addition, in jets themselves, as well as in cooler gases, magnetic pressure support can be important. Magnetic fields can affect the development of thermal instabilities \citep{2018MNRAS.476..852J}, as well as the kinematics and the evolution of the cool filaments in massive systems, which may in turn impact the feeding and feedback of the SMBH \citep{2021MNRAS.504..898W, 2022MNRAS.510.3778M}. 

Over the past decade, cosmic ray feedback has also been studied extensively in galaxy simulations. In almost all galaxy-scale simulations, cosmic rays are modeled as a separate fluid with a different equation of state. It can interact with the thermal fluid via streaming and/or diffusion. Cosmic rays do not suffer from radiative cooling loss as gas does. Moreover, additional cosmic ray pressure support can reduce the density of the gas, and therefore its cooling rate. Thus cosmic rays can make feedback more efficient. That is, cooling flows can be suppressed with a lower total $\dot{E}_{\rm BH}$ when some of the energy is injected into cosmic rays.

In the ICM, observations suggest that cosmic ray pressure is small, which is very different from the interstellar medium (ISM) where cosmic rays reach equipartition with gas and magnetic fields. Nevertheless, simulations have shown that cosmic ray heating can provide efficient feedback within the observational constraints \citep{2008MNRAS.384..251G, 2017MNRAS.467.1449J, 2017ApJ...844...13R, 2021MNRAS.507..175S}. In addition, cosmic rays can also affect the development of thermal instability in the hot phase \citep{2019ApJ...871....6Y, 2020ApJ...903...77B}, which can then impact the duty cycle of AGN feedback.

\subsection{Understanding AGN feedback in simulations}\label{sec_sim_heating}

Observationally, we can only capture single ``snapshots'' of AGN feedback, or a short period ($\sim$ years) compared with timescales relevant for galaxies. Our understanding of AGN feedback is also limited by the spatial and spectral resolutions of the observational data, as well as the limited wavelength coverage. On the other hand, it is easier (albeit only in a relative sense) to analyze numerical simulations which are only limited by the numerical spatial and temporal resolutions. However, analyzing numerical simulations can also be rather challenging because of the degeneracy of different aspects of sub-grid models, as well as the degeneracy between numerical effects and physical effects. In addition, a simplistic numerical model is usually not a good representation of reality, while in a model with more physical processes, the different components (e.g., AGN vs star formation) can have complex interactions which are difficult to understand.

Over the past decade or so, many numerical models of AGN feedback have achieved some level of success in ``quenching'' massive galaxies and suppressing cooling flows. In recent years, efforts have been made to understand exactly how AGN does it in these models, and how we can connect models to the observations. Here, we mainly focus on idealized simulations of isolated massive galaxies or galaxy clusters.

Before discussing successful models, it is helpful to first review the failed models. As is discussed earlier, two types of models have generally failed regardless of the feedback efficiency $\epsilon$: simple thermal energy injection and narrow fast jets. Simple thermal feedback usually results in energy loss through radiation, while narrow fast hydro jets dump energy outside of where it is needed. 

A successful model needs to deliver the energy to the right place without wasting too much of it to radiation. This is why most successful models include some form of kinetic feedback that is able to cover a large fraction of the host's central region. Depending on the exact implementation of the model, how energy is deposited into the ICM may differ.

In models where a large amount of thermal energy is injected with the jets, turbulent mixing of the hot jet material and the ICM can be a major heating mechanism \citep{Hillel2016}. In simulations with mostly kinetic jets or winds, the kinetic energy can be thermalized via shock waves, sound waves, and turbulent dissipation. Recent analyses suggest that although shock waves are not volume-filling in simulations, they are the most important way to thermalize kinetic energy carried by the jets \citep{2017ApJ...847..106L, 2022arXiv220105298W}. Sound waves can be important but are subdominant, especially in a more realistic setup with multiple AGN outbursts \citep{2019ApJ...886...78B, 2022arXiv220105298W}. The level of turbulence is found to be rather low in most AGN feedback simulations, including simple plane-parallel models \citep{2015ApJ...815...41R, 2018ApJ...857...84B} and more realistic setups \citep{Hillel2016, 2016ApJ...829...90Y, 2017ApJ...847..106L, 2017MNRAS.470.4530W, 2017MNRAS.472.4707B, 2018ApJ...863...62P}. 

The low level turbulence in simulations appears to conflict with Hitomi observations \citep{2016Natur.535..117H} and Chandra X-ray surface brightness fluctuation analysis, which suggest a level of turbulent dissipation enough to balance radiative cooling \citep{2014Natur.515...85Z}. This tension has motivated further studies from both the theoretical side and the observational side. The investigation is far from conclusive, but ideas have been proposed, including alternative interpretations of the Hitomi and Chandra observations \citep{2017MNRAS.464L...1F}. In most numerical simulations, the velocity dispersion on tens of kpc scales is usually in good agreement with the Hitomi results \citep[e.g.,][]{2018ApJ...854..167G, 2018ApJ...863...62P} even when turbulent dissipation is a subdominant heating mechanism \citep{2017ApJ...847..106L}. Most of the bulk motion in simulations does not seem to cascade to smaller scales as turbulence. \citep{2020ApJ...889L...1L} analyze the kinematics of the cooler filaments and find that the level of turbulence on small scales is indeed lower than previous estimations assuming Kolmogorov cascade. If the cooler filaments are reliable tracers of the hot ICM, then ICM turbulence has a spectrum steeper than Kolmogorov. The exact reason for this is a subject of debate and may be related to magnetic fields, shocks, and/or gravity waves \citep{2021MNRAS.504..898W, 2022MNRAS.510.2327M, 2022arXiv220304977H, 2022arXiv220304259Z}.

In addition to generating heat, kinetic jets can provide feedback in another way: by doing work on the surrounding gas and uplifting it to larger radii in galaxy clusters, or outside the galaxy in smaller systems. This process does not compensate for the radiative cooling loss directly, but can very effectively suppress or delay further cooling. The amount of energy involved in adiabatic processes can be rather significant in simulations with jet feedback and is crucial in successfully quenching galaxies \citep{2016ApJ...829...90Y, 2017ApJ...847..106L, 2017MNRAS.470.4530W}. Somewhat similarly, the main role of cosmic rays in feedback is also reducing cooling rather than generating heat. Yet cosmic ray feedback can be very efficient. Recent jet simulations with cosmic rays are able to achieve quenching with order-of-magnitude lower energetics \citep{2021MNRAS.507..175S}.


The above summarizes our exploration of the ``quenching'' or ``heating'' effect of AGN feedback, with the main result being a suppression of star formation. It is therefore sometimes called ``negative feedback'' (see also Section~\ref{sec:models}). It has been shown that AGN feedback can enhance star formation in certain situations, and the effect is often called ``positive feedback''. The positive AGN feedback in galaxies usually refers to the compression of multiphase ISM which can trigger star formation \citep{2013ApJ...772..112S}. In galaxy clusters, positive feedback refers to the formation of the multiphase filaments as a result of AGN feedback. 

Most galaxy clusters are observed to have a $t_{\rm cool}$ longer than $10t_{\rm dyn}$ and the ICM should be thermally stable \citep{1989ApJ...341..611B}. Yet many cool-core galaxy clusters and groups harbor multiphase filaments in the center. These filaments produce bright $H_\alpha$ as well as enhanced soft X-ray emissions. Some of the filaments are also observed to have molecular gas and form young stars. 

\begin{figure}[!ht] 
\centering
\includegraphics[width=0.88\columnwidth]{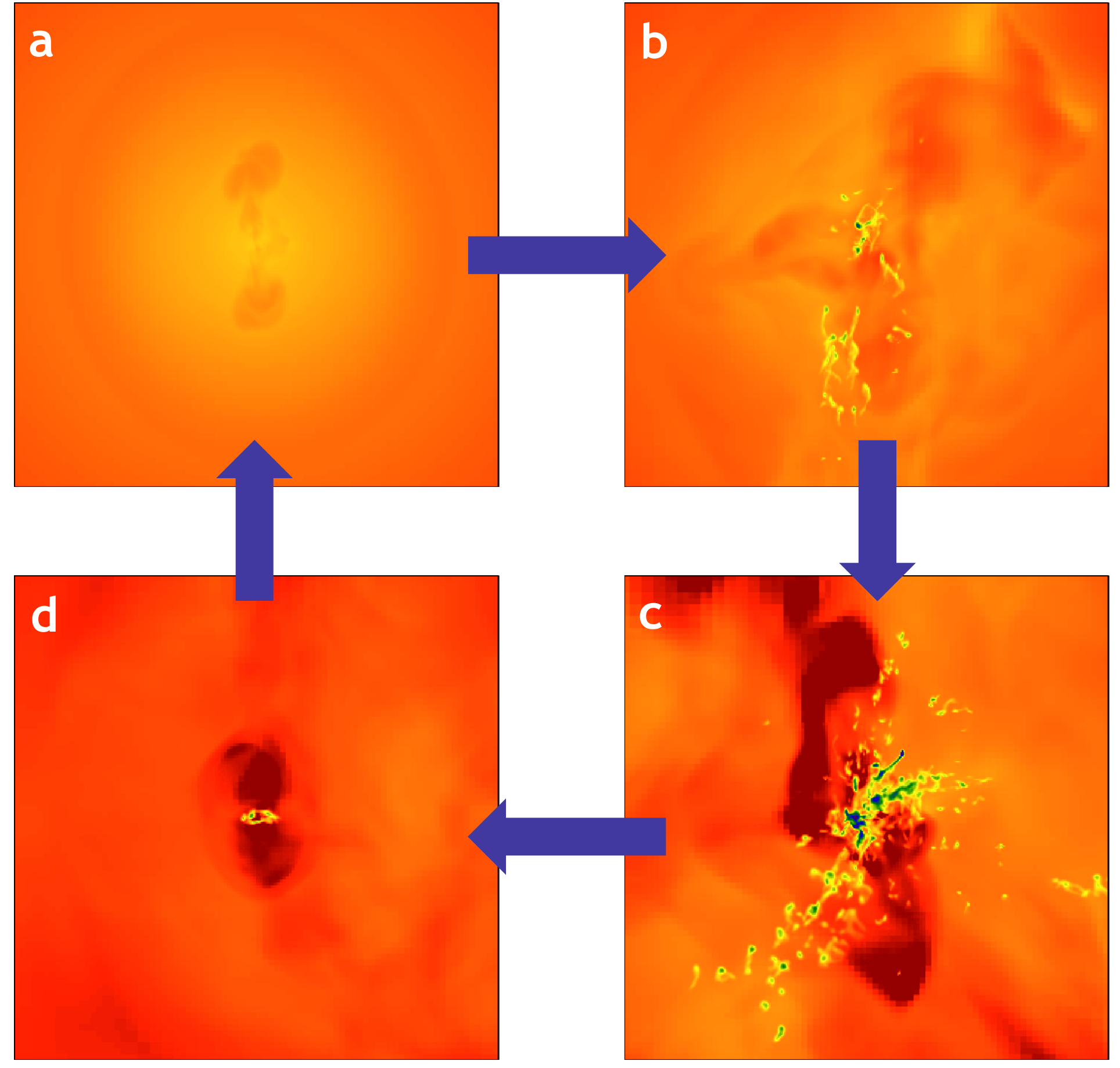}
\caption{An AGN feedback cycle ($\sim$ Gyr) based on an idealized simulation of a Perseus-like cluster with jet feedback \citep{2015ApJ...811...73L}. The size of an individual frame is 80 kpc across, and color corresponds to the projected temperature of the ICM (red is hot and blue is cold). Panel a shows the beginning of the cycle when the onset of a cooling flow powers AGN jet feedback. Jets then trigger thermal instability in the ICM (panel b). The accretion of the cold gas boosts AGN power, which stimulates more cooling and star formation (panel c). The strong feedback heats up the ICM, which slows down further condensation and star formation (panel d). Reduced condensation results in reduced SMBH accretion, and thus reduced heating. This allows the ICM to return to an asymptotic cooling flow solution (back to panel a and the beginning of the next cycle).}
\label{fig:feedback_loop}
\end{figure}

Idealized simulations have shown that local thermal instability can develop in a stratified atmosphere where cooling is globally balanced by volumetric heating \citep{2012MNRAS.419.3319M, 2012MNRAS.420.3174S}. Simulations with buoyant bubbles show that cooling can proceed in the wake of the rising bubbles \citep{2008A&A...477L..33R}. In simulations with jet feedback, the kinetic jets can uplift low entropy gas to trigger thermal instability \citep{2014ApJ...789..153L}. Precipitation can also be stimulated by turbulence driven by AGN feedback \citep{2018ApJ...868..102V}. Recent simulations that include both kinetic and radiation feedback show that cool filaments can form out of warm ionized outflows from the AGN \citep{2020NatAs...4..900Q}.

Overall, numerical simulations suggest that the formation of cool filaments is tightly linked to AGN feedback in galaxy clusters. Since some of the cool filaments can form stars, AGN triggered cooling in clusters can be viewed as positive feedback. However, the main global effect of AGN feedback is still heating rather than cooling. The interplay between heating, cooling, and star formation can result in an AGN feedback cycle illustrated in Figure~\ref{fig:feedback_loop}. 
Note that such a cycle can be interrupted or ``contaminated'' by other processes in reality. For example, mergers can also trigger AGN outbursts. In galaxy groups and massive galaxies, some cool gas may be directly brought in by satellite galaxies or accretion of cold streams from cosmic filaments.

\subsection{Modeling SMBH accretion in simulations}\label{accretionYuan}

The process of SMBH feeding is tightly linked to its feedback. Similar to feedback, due to the resolution limit, galaxy-scale or cosmological simulations rely on sub-grid models for SMBH accretion. 

Most cosmological simulations use the Bondi rate with the maximum limited by the Eddington rate. When calculating the Bondi rate, some simulations use the local gas property very close to the SMBH (e.g., Illustris) while some use a kernel-weighted average over a number of neighboring cells (e.g., TNG). Many simulations include a boost factor $\alpha$ to compensate for the unresolved multiphase ISM and to boost the accretion rate onto SMBHs. Some simulations use a simple constant $\alpha$ (often set to be 100) while in some, $\alpha$ can be more complicated. For example, Horizon-AGN uses the formula in \citep{2009MNRAS.398...53B} where $\alpha$ is a function of the local gas density and is only boosted when the density is above a certain threshold. A similar scheme is also used in the Magneticum simulations and the BAHAMAS project.

In idealized simulations, especially in those that can resolve the Bondi radius, a different accretion scheme can be used which takes the flux through the inner boundary as the accretion rate \citep[e.g.,][]{1997ApJ...487L.105C, 2018ApJ...864....6Y}. Another similar accretion scheme often used is to calculate the accretion rate using the cold gas mass in the close vicinity of the SMBH divided by the dynamical time \citep[e.g.,][]{2005ApJ...632..821P, 2014ApJ...789..153L, 2016ApJ...829...90Y, 2017MNRAS.471.1531P, 2017ApJ...841..133M}. This is sometimes termed cold accretion (as opposed to the hot Bondi accretion). The two schemes can give similar results when multiphase gas is present, because the accretion is usually dominated by the cold phase, and the rate is much higher than the Bondi rate evaluated at the Bondi radius
\citep{2013MNRAS.432.3401G, 2019ApJ...877...47Q}.

Whether the hot or the cold mode accretion dominates is still a subject of debate, and the answer likely depends on the system the SMBH is in. Although SMBH accretion is an important astrophysical problem, from the modeling perspective, it appears that the simulation results are not very sensitive to the accretion model. For example, \citet{2017ApJ...841..133M} show that with reasonably high resolution, simulations with different accretion schemes have similar results. The general success of modern cosmological simulations also suggests that simple (boosted) Bondi schemes may be sufficient in large-scale simulations. When comparing across different simulations, it is usually the feedback schemes that determine the simulation outcome \citep[e.g.,][]{2022MNRAS.509.3015H}.


\section{Conclusion}
Groups and clusters of galaxies represent the most massive end of virialized objects in the Universe. They possess hot gaseous atmospheres with the characteristic temperature in the keV range. Giant elliptical galaxies reside in their centers and typically host a few billion solar masses black holes.

Radiative energy losses of the gas in the cores of clusters (if not compensated by a powerful source of energy) should lead to catastrophic cooling and massive star formation in stark contrast with observations. A consensus has emerged over the last two decades, that the activity of central supermassive black holes provides the needed energy. Energetics arguments suggest that these black holes can potentially release enough energy, but the efficiency of the released energy conversion into gas heating must be very high. From observations we know that most of these black holes appear in the vast majority as Fanaroff-Riley type I radio sources and can be classified as LERG based on their optical properties. This suggests that these black holes are accreting matter at a rate significantly smaller than the Eddington rate, placing them in the regime of radiatively inefficient accretion, when most of the energy is released in a mechanical form. Almost independent on the detailed physical picture, the conversion efficiency of the mechanical energy into gas heating is very high (close to 100 per cent) opening the possibility for sub-Eddington accretors to act as the energy source for the cooling gas. It is still debated if the black holes are fed directly by the hot gas (hot mode or Bondi-type accretion) or by the gas that first cools and then gets accreted (cold mode).

Such radio mode AGN feedback in clusters and groups is different from the feedback in less massive systems in several important aspects. For instance, it is not capable of expelling or dramatically affecting all baryons in a cluster, but is able to maintain the gas temperature in the core in the keV range. For this reason, radio mode AGN feedback is sometimes called maintenance mode. 

The dominant form of mechanical energy released by AGN and its intermittency are also actively debated, including jets, winds/outflow, which differ in the degree of collimation and amount of momentum they carry. One of the popular models postulates that jets quasi-continuously inflate bubbles filled with relativistic particles in the ICM, so that the enthalpy of the bubbles accounts for the largest fraction of the released energy. This assumption provides an order of magnitude estimates of the AGN power across the entire range from small  groups to large clusters, which match approximately the cooling losses of their host clusters. This on the one hand suggests that these models are broadly correct and on the other hand, that AGN adjust their power to achieve an approximate balance between cooling and heating.

Several models for ``transporting'' mechanical energy from supermassive black holes to the gas have been suggested, including buoyantly rising bubbles, mixing, strong shocks, sounds waves, cosmic ray streaming to name a few. While it is still debated which process dominates, all of them are very efficient - a distinct feature of the mechanical feedback. 

In terms of observations, it has now become clear that the central AGN in groups and clusters of galaxies do much more than compensate radiative losses of the cooling gas. The radio jets also appear to transport metals out to dozens and even hundreds of kpc. They also appear to - systematically - be able to drive massive molecular outflows out to dozens of kpc, indicating that quasars are not the only ones that are able to create such outflows. Such feedback also appears to be operating over several Giga-years, implying that radio mode feedback is a dominant part of the lives of groups and clusters. 

Yet another actively debated question is the origin of the cool gas ($T\lesssim 10^4\,K$) in cluster cores, which is often observed as filaments bright in e.g. $H_\alpha$ line. The options discussed include recent cooling from the hot phase, or recycling of the gas that has cooled long time ago and then lives in the system powered by some (presently not well understood) mechanism.  

Overall, there is a consensus that mechanical AGN feedback plays a fundamental (maybe even dominant) role in the cores of clusters and groups and new observational data from existing and forthcoming facilities will be able to pinpoint particular physical mechanisms behind the gas heating.     


\bibliography{references} 

\end{document}